\documentclass[showpacs,aps,prb,twocolumn,groupedaddress]{revtex4-2}

\usepackage[latin1]{inputenc}
\usepackage{graphicx}
\usepackage{textcomp}
\begin{document}

%\preprint{}
\title{Magnetic moment orientation and in depth distribution of dysprosium near the surface of DyCo$_{4.6}$ thin films determined by the analysis of their X ray circularly polarized absorption Dy $M_{5}$ spectra}


\author{J. D\'iaz,$^{1,2}$}
\email[]{jidiaz@uniovi.es}
\author{C. Blanco-Rold\'an,$^{1,2}$}
%\author{C. Blanco,$^{1,2}$ C. Quir\'os,$^{1,2}$ S. M. Valvidares,$^{3}$ J. M. Alameda,$^{1,2}$ }
\affiliation{$^{1}$Universidad de Oviedo, Calle Federico Garc\'ia Lorca, 18, Oviedo 33007, Spain}
\affiliation{$^{2}$CINN (CSIC-Universidad de Oviedo), 33940 El Entrego, Spain}


\date{\today}

\begin{abstract}

We have investigated the RE atomic distribution and its magnetic moment orientation at the region near the surface of DyCo$_{4.4 \& 4.6}$ ferrimagnetic amorphous films with perpendicular magnetic anisotropy (PMA). XMCD spectroscopy of the films at the Dy $M_{4,5}$ and Co $L_{2,3}$ edges using total electron yield (TEY) detection was performed at 2 K and 300 K temperatures, and at sample orientations ranged from 0$^{\circ}$ to 70$^{\circ}$ with respect to the magnetic easy axis. The measurements showed an apparent partial decoupling between the cobalt and dysprosium magnetic sublattices. At RT, the magnetic moment per atom of dysprosium was below the minimum value expected if all dysprosium moments were AF coupled to cobalt. At 2 K, the cobalt sublattice presented a stronger magnetic anisotropic behavior than the dysprosium sublattice. A detailed analysis of the circularly polarized spectra of the Dy $M_{5}$ edge, based on the deconvolution of the spectra in their related parallel, antiparallel and transverse to $J_{z}$ spectral components, demonstrates that the spectra are composed by dysprosium with different magnetic moment distributions. The fit of the Dy $M_{5}$ spectra using the $J_{z}$ spectral components evidenced a gradation of dysprosium concentration due to segregation at the region probed by TEY. The topmost layer was magnetically uncoupled from cobalt. At RT, 25$\%$ of the dysprosium magnetic moments in the under layer were found averaged oriented in the same direction as cobalt. The expected weak magnetic coupling of these dysprosium atoms to cobalt should explain the surprisingly lower magnetic anisotropy of the dysprosium sublattice compared to that of cobalt probed by TEY at 2 K.

\end{abstract}

%\keywords{}
\pacs{75.30.Gw, 75.50.Kj,75.70.Ak, 78.70.Dm,81.05.Bx}

\maketitle

\section{Introduction}
RE-TM alloys are well known magnetic materials since decades ago \cite{buschow_RMreview77,SKOMSKI_AniRE}, constituting  a key component in a variety of industrial and technological applications. A renewed interest in these materials has grown  because their richness in magnetic behaviors and their relative flexibility for tailoring their properties to fit in specific magnetic applications and devices. RE-TM alloys are present in spring magnets\cite{GdCo_Cristina, GdCo_Rafa,SciRep_2015}, magnetic topological formations \cite{CrisNat_2015,GdCo_Skyrmions, AureNat_2020}, spin wave functional devices \cite{Luis_APL} and all optical magnetic switching \cite{2014_enginering_nature, 2014_All_Optical_APL, SpinFlop_PRL}. Their use is also favored by its simple thin film preparation process which can be done at RT. 

The extraordinary magnetic properties of these alloys are based on the strength of the TM-TM exchange coupling, the high magnetic moment of the RE, and their RE-TM indirect exchange coupling interaction. The high orbital moment of its unquenched $4f$ orbital and the high energy of its spin-orbit coupling, makes the RE acts as a strong and localized magnetic moment whose orientation depends on the interatomic exchange and the crystal field at its particular local atomic environment. These two interactions are mainly provided by their TM neighbors whose RE-TM interaction energies overwhelms those of RE-RE. Since the RE-TM magnetic exchange coupling is antiferromagnetic in spin, RE-TM alloys are ferrimagnetic for heavy REs like dysprosium. This interaction is usually treated as a molecular field whose intensity is calculated to be of the order of 200 T for crystalline DyCo$_{5}$ \cite{MolecularField_DyinCo}. The energy of this interaction is, at least, one order of magnitude smaller than the TM-TM exchange. These range of values are comparable to $k_{B}T$ energies, giving rise to different magnetic configurations as a function of temperature and RE concentration. A characteristic parameter that define these ferrimagnetic alloys is their compensation temperature, $T_{Comp}$, the temperature where the magnetic moment of the RE and TM sublattices cancel to each other. 

Giving the radical different magnetic behavior of the RE and the TM, the understanding of the magnetism of these alloys requires a precise characterization of the magnetic moment and anisotropy of the RE and TM sublattices separately. The perfect technique to do so is XMCD spectroscopy. The magnetic properties of the RE and TM sublattices can be studied separately by tuning the incident circularly polarized X rays to the corresponding absorption edges of the RE ($M_{4,5}$, probing its $4f$ orbital) and the TM ($L_{2,3}$, probing its 3d orbital)\cite{Stohr_JM3review,XMCD_US,teramura_RE}. This is actually the technique of choice for the study of RE-TM alloys in the form of thin films because it allows ex-situ sample preparation and it can be sensitive to regions at different depths of the sample by changing the way the  X ray absorption spectra is detected. Surface sensitivity is attained using TEY detection, with probed depths of the order of 2 to 3 nm, whereas fluorescence yield or X ray transmission are more bulk sensitive techniques. The simultaneous use of these two detection modes has been proved to be important for a correct understanding of these alloys. The different $T_{Comp}$ measured in these alloys using TEY and bulk detection\cite{SciRep_2015}, together with the reported RE segregation at their surface\cite{GdCo_segre_PRB,TbFeCo_segre_PRM}, has been used as an argument to explain interesting phenomena that occurs at temperatures near their $T_{Comp}$ when the applied magnetic field is intense enough. The hysteresis loops of some DyCo alloys of similar concentration than the samples used in this study presented side wing loops at high applied fields as the presented in some spring magnet-like bilayer structures \cite{Givord,GdCo_Cristina, GdCo_Rafa}. The effect was explained as derived from the effective different RE concentration at the bulk and the region near the surface which will yield two different $T_{Comp}$ for each region \cite{SciRep_2015,SciRep_2019}.

In these studies, the $T_{Comp}$ at the region near the surface was estimated from the magnetic moments of cobalt and dysprosium deduced by XMCD measurements done using TEY detection.  In all the cases, the analysis have always assumed the presence of a single dysprosium magnetic phase. However, due to the structural disorder of these alloys, it is expected different grades of intensity in the magnetic exchange interaction between dysprosium and cobalt. Such effects should be stronger at the surface whose assumed larger RE concentration is driven by RE-RE bonding preference. Previous XMCD studies in NdCo alloys conducted by us using TEY detection \cite{NdCo_PRB} reported a substantial proportion of RE atoms that behaved as a paramagnetic, indicating that not all the RE atoms probed might have the same exchange interaction strength with the TM and, therefore, the same magnetic orientation distribution. Although those experiments were sensitive to the surface, their exact distribution was not demonstrated. Actually, EXAFS experiments performed by us in similar samples deduced the possible presence of RE segregation in the bulk as well\cite{NdCo_EXAFS}. The possible existence of paramagnetic RE atoms non-exchange coupled to the TM  due to disorder or/and RE segregated would contribute to overestimate the $T_{Comp}$ at the surface because its average magnetic moment orientation would be in the same direction of the appied field. 

This effective bilayer interpretation of the side wing loops in DyCo monolayers has been contested by others who consider that the observed effect is due to a spin flop phase transition\cite{HT_phaseDiagram}. The range of temperatures and fields under which this transition is produced can be predicted by the $H-T$ (applied field $H$-Temperature) phase diagram of the alloy which is built by considering all the interactions present in the alloy. These studies shows that the relative magnetic anisotropy of the RE and TM sublattices \cite{HT_phaseDiagram}, and the interaction of the alloy at the interface with other metals \cite{HT_phaseDiagram} are key to understand these spin flop transitions. 

It is clear, then, the importance of having a good characterization of the magnetism of these alloys at the region near the surface or at their interface with other materials whose magnetic structure could be complex. The purpose of the experiment presented in this work is to improve the spectroscopical tools to understand the magnetic behavior of the TM and the RE sublattices, their mutual interaction and their magnetic anisotropy using the information extracted from the X ray absorption spectra of the RE atoms, and considering their possible inhomogeneous distribution in depth at the region probed by TEY.

For this experiment, we prepared DyCo thin films anisotropically uniaxial with high PMA energies.  Their anisotropy fields at 2 K, $H_{K}$, were well above the range of available field intensity in the experiment. Measurements were done at 2 K and RT. These two temperatures were distant enough from the $T_{Comp}$ of the alloys to avoid possible spin flip (and flop) effects. Also, it permitted to measure in the regions where each of the sublattices were magnetically dominant (cobalt at RT, and dysprosium at 2 K). At 2 K, thermal disorder was reduced to a minimum. 

The ferrimagnetic character of DyCo alloys permits to detect dysprosium atoms magnetically uncoupled from cobalt since they will be oriented in the opposite direction than their counterpart exchange coupled dysprosium atoms at temperatures above $T_{Comp}$ (cobalt sublattice magnetically dominant). Their detection requires a deconvolution of the RE $M_{4,5}$ spectra in their parallel, antiparallel and transverse components of $J_{z}$ which are specially well defined in dysprosium. This deconvolution has been presented before in DyCo films  \cite{deconvoluted_DyCo}, but it was not linked to the moment orientation of the dysprosium magnetic moments as it is done and explained in detail in this work. Also, we use the XMCD spectra to deconvolve these components from the circularly polarized spectra, instead of relying completely on their theoretically calculated shapes, as the mentioned work tried. Actually, this is the first time that this kind of analysis is applied in a certainly complex system like the studied in this work.

%We show the potential of this novel technique in the study of the interaction between RE and cobalt atoms at the region probed by TEY, which is used to detect and characterize the presence of a RE segregated layer. The total magnetic anisotropy of the alloy should depend on the distribution of the crystal field orientation at the RE sites\cite{SKOMSKI_AniRE}, and its influence on the magnetic moment of the bonding TMs at each RE site. Following the single ion anisotropy model \cite{SKOMSKI_AniRE}, both crystal field and RE-TM indirect exchange are expected to be tightly related. Therefore, the RE environments that most contribute to the magnetic anisotropy of the alloys should be strongly exchange coupled to the TM as well. A way to characterize and even deduce the RE anisotropic environment distribution, and the ratio between anisotropy energy and exchange interaction is to perform magnetometry measurements at different field orientations with respect to the easy axis of the sample in the non saturation regime, in a similar way that it is done, for instance, in magnets formed by disordered oriented nanomagnets \cite{Honorino_APL}, providing that the magnetization of the sample rotates without breaking in magnetic domains.

We show the potential of this novel technique in the study of the interaction between RE and cobalt atoms at the region probed by TEY, by measuring at different orientation angles under a strong PMA. This technique is able to detect the presence of a RE segregated layer that affects in a significant way the measured moment of the RE. The presence of this layer seems intrinsic to the thin film growth in RE-TM alloys. Even removing the effect of this layer, the proportion of dysprosium atoms directly engaged in the PMA anisotropy of the alloy probed by TEY was not majority, possibly  due to the extended thickness of the cobalt depleted layer caused by the dysprosium segregation.  

The paper is organized as follows:
First, sample preparation and experimental details for XMCD data acquisition are described. Next, after showing the VSM magnetometry characterization of the measured films, the XMCD experimental results for each elemental sublattice are presented and discussed separately. The following section is dedicated to explain how the deconvolution of the Dy $M_{5}$ spectra is made, with the calculation details shown in an Appendix section. After this explanation, a model to fit the Dy $M_{5}$ of the measured samples using the deconvolved spectral components is proposed, which consisted in two layer with different dysprosium moment distributions. The results of the fits are presented and discussed. The final section are the conclusions of the work.

\section{Experiment \label{experiment}}
The two studied DyCo thin films were prepared at RT by magnetron sputtering at a base pressure of $10^{-8}$ mbar and $10^{-3}$ mbar Ar pressure. They were grown onto silicon wafer substrates using two separate magnetron guns set at normal (cobalt gun) and at 30$^{\circ}$ angle incidence (dysprosium) with respect to the normal to the sample. Dysprosium and cobalt concentrations were calibrated using a quartz balance. The deposition method was different for each sample. Sample called DCC was prepared by codeposition of cobalt and dysprosium. Sample called DCM was grown by the alternate deposition of cobalt and dysprosium layers. The nominal thickness of the cobalt and dysprosium layers was 4.9 \AA\ and 2.8 \AA\ respectively. The topmost deposited layer was cobalt. This second preparation method was intended as a way to estimate the importance of the interdiffusion between the elements forming the alloy in their structure and magnetic properties. All the samples were protected with a 20 \AA\  thick aluminum capping layer. The mean atomic concentration in the alloys was determined by electron induced fluorescence spectroscopy. The difference in concentration between both films was relatively small. Sample DCC contained more dysprosium (DyCo$_{4.4 \pm 0.05}$)  than the DCM film (DyCo$_{4.6\pm 0.05}$). Their nominal thickness was 35 nm.

X ray circularly polarized absorption spectra were obtained at the HECTOR endstation \cite{HECTOR} of the BOREAS BL-29 beamline at the ALBA synchrotron using total electron yield detection. HECTOR has a cryomagnet that can apply up to  $\pm$6 T along the x-ray beam direction at different sample orientations, which in our case ranged from normal incidence (0$^{\circ}$) to near grazing incidence (70$^{\circ}$). The cryomagnet works at ultra-high vacuum conditions (pressure within the 10$^{-10}$ mbar range). A liquid He cryostat permits to fix the sample temperature between 2 K and 350 K. Circularly polarized light was produced by an APPLE II elliptical undulator. Each XMCD spectra was the result of 4 spectra taken at opposite circular polarization helicities and magnetic field orientations. 

\section{EXPERIMENTAL RESULTS and DISCUSSION}

\subsection{Magnetometry \label{magnetometry}}
Both samples presented a PMA even at RT. Table \ref{tab_Comp_Temp_DCC_DCM} shows the magnetization in remanence of samples DCC and DCM measured using a Vibrating Sample Magnetometer (VSM). The values were obtained after magnetic saturation of the films with a field of 9 T at 5 K applied normal to their plane \cite{Tesis_Cristina}. The same table gives the compensation temperatures, $T_{Comp}$, of the alloys and their coercive fields at 2 K and RT. $T_{Comp}$ temperature was smaller in sample DCM (90 K) than in sample DCC (120 K), in agreement with their different RE concentration, and with the expected by comparing with the reported by others in DyCo alloys of similar concentrations \cite{SciRep_2015,SciRep_2019}, assuming a linear relationship between $T_{Comp}$ and the atomic concentration of the alloy \cite{TbFeCo_segre_PRM}. The higher cobalt concentration in sample DCM causes a marked lower magnetic remanence at 10 K than in sample DCC. This explains the large difference between the coecive fields measured at 10 K, higher in sample DCM (3.5 T) than in sample DCC (1.9 T). However, sample DCM has a higher $H_{C}$ at RT, when its $M_{S}$ is higher. X ray reflectometry shows rougher surfaces in DCM than in DCC thin films, suggesting that the larger $H_{C}$ of the DCM thin film is probably caused by its higher density of domain wall pinning defects. The coercive fields of both samples were notoriously higher than the reported in DyCo$_{4}$ by others \cite{SciRep_2015,SciRep_2019}, with $H_{C}$ values lower than 1 T at a similar decrement ($\Delta$T=150 K) in temperature from their $T_{Comp}$ (250 K). This might account for a higher PMA energy and/or domain wall pinning defects density in our samples. None of the samples presented triple hysteresis loops at fields below 6 T at temperatures close to their $T_{Comp}$ as the reported in samples of similar concentration and thickness \cite{SciRep_2015,SciRep_2019}.

\begin{table}[h]
\caption{Compensation temperature, $T_{Comp}$, Magnetic remanence, $M_{R}$, and coercive field, $H_{C}$, of samples DCC and DCM at 2 K and RT\cite{Tesis_Cristina}.\label{tab_Comp_Temp_DCC_DCM}}
\begin{ruledtabular}
\begin{tabular}{lllllll}
Sample & $T_{Comp}$&Temperature  & $M_{R}$ (emu$\cdot$cm$^{-3}$)& $H_{C}$ (T)\\	%\cline{7,7}
\colrule\hline
DCC&90 K& 2 K&125& 1.9 \\
&& 300 K&250& 0.2 \\
DCM&125 K& 2 K&75& 3.5 \\
&& 300 K&275& 0.4 \\
 
\end{tabular}
\end{ruledtabular}
\end{table}

\subsection{XMCD hysteresis loops and magnetic moments \label{XMCD_hystLoops}}

\begin{figure}
\includegraphics[width=8 cm]{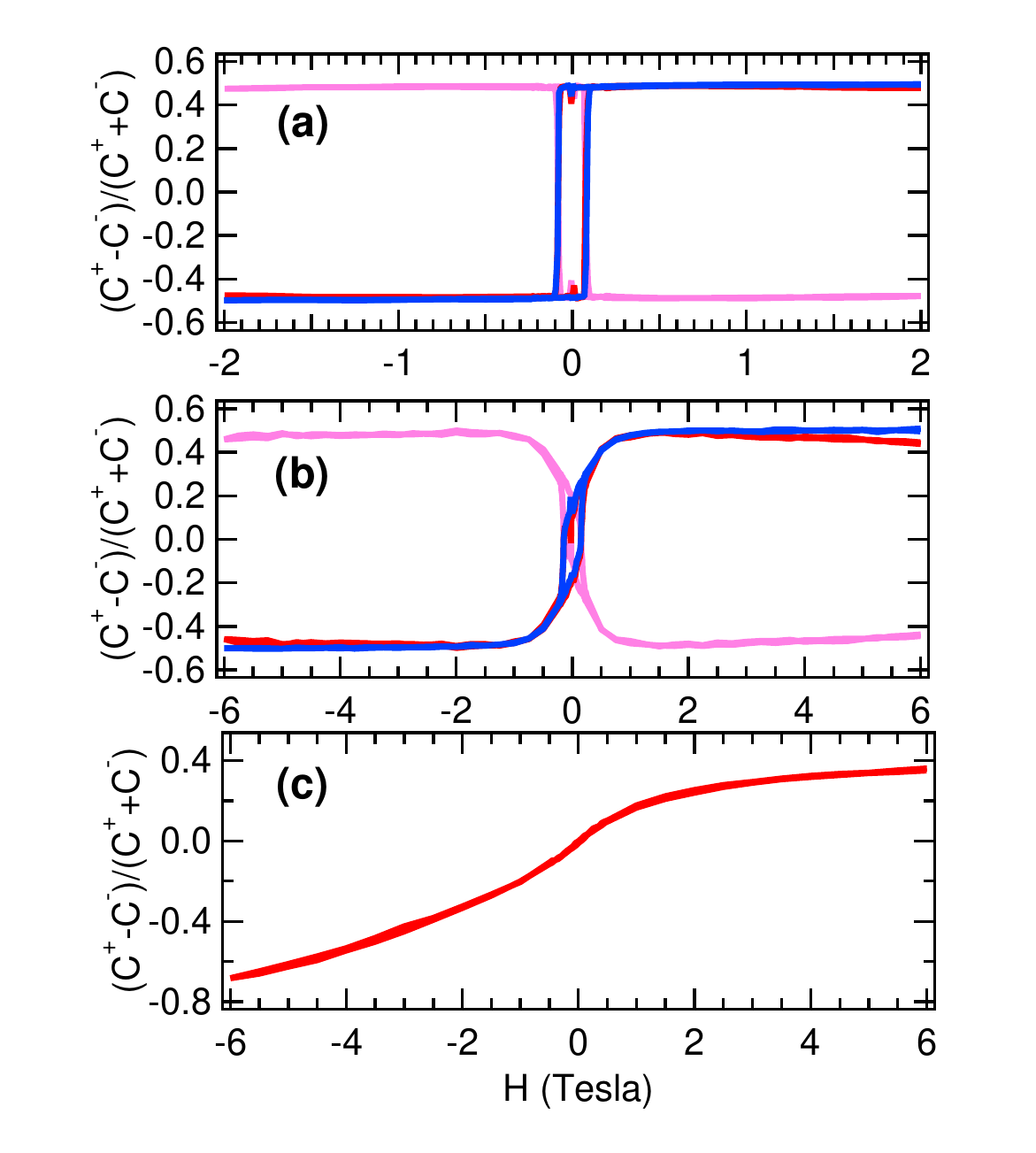}
\caption{Hysteresis loops taken at the Dy $M_{5}$ (pink) and Co $L_{3}$ (blue) edges at RT: (a) normal orientation, (b) grazing incidence (70$^{\circ}$). The hysteresis loop of the dysprosium sublattice multiplied by -1 has been drawn in red behind the cobalt loop for comparison. (c) Dysprosium hysteresis loop taken at 2 K at grazing incidence (70$^{\circ}$). \label{Fig_XMCD_Loops}}
\end{figure}

Figure \ref{Fig_XMCD_Loops} shows the hysteresis loops of sample DCC at RT and 2K for the cobalt and dysprosium sublattices obtained by measuring the intensity of the Co $L_{3}$ peak and the Dy $M_{5}$ peak respectively. Their change in shape with field orientation at RT (figures \ref{Fig_XMCD_Loops} (a) and (b)), from squared at 0$^{\circ}$ orientation to "S" shaped at 70$^{\circ}$ field orientation, shows that the samples had a PMA, as observed by VSM magnetometry. The loops of the cobalt and dysprosium sublattices are nearly identical. A small decoupling between both sublattices is only noticeable at high fields. The magnetization of cobalt remains constant at high fields up to 6 T whereas that of dysprosium seems to decreases steadily with the increasing field from zero field. 

A similar effect, although much more pronounced than in the present case, has been observed in NdCo films \cite{NdCo_PRB}. The effect was demonstrated to be caused by a portion of paramagnetic Nd. A similar explanation would justify the observation in the present case since, at this temperature, any dysprosium magnetically uncoupled to cobalt would be oriented towards the field direction as cobalt, reducing its total magnetization.  

At 2 K, only the hysteresis loop of the dysprosium sublattice in sample DCM was taken at near the plane field orientation (70$^{\circ}$), shown in figure \ref{Fig_XMCD_Loops}(c). The shape of the loop shows that at 6 T the films were far from being magnetically saturated, indicating that the PMA energy of the samples was strongly increased at this temperature. The lack of coercive field in the loop shows that the measured magnetic moment experience a progressive rotation with the applied field intensity. The loop shows a shape asymmetry between the positive and negative branches.

\subsection{XMCD Cobalt\label{Co spectra}}
\begin{figure}
\includegraphics[width=8 cm]{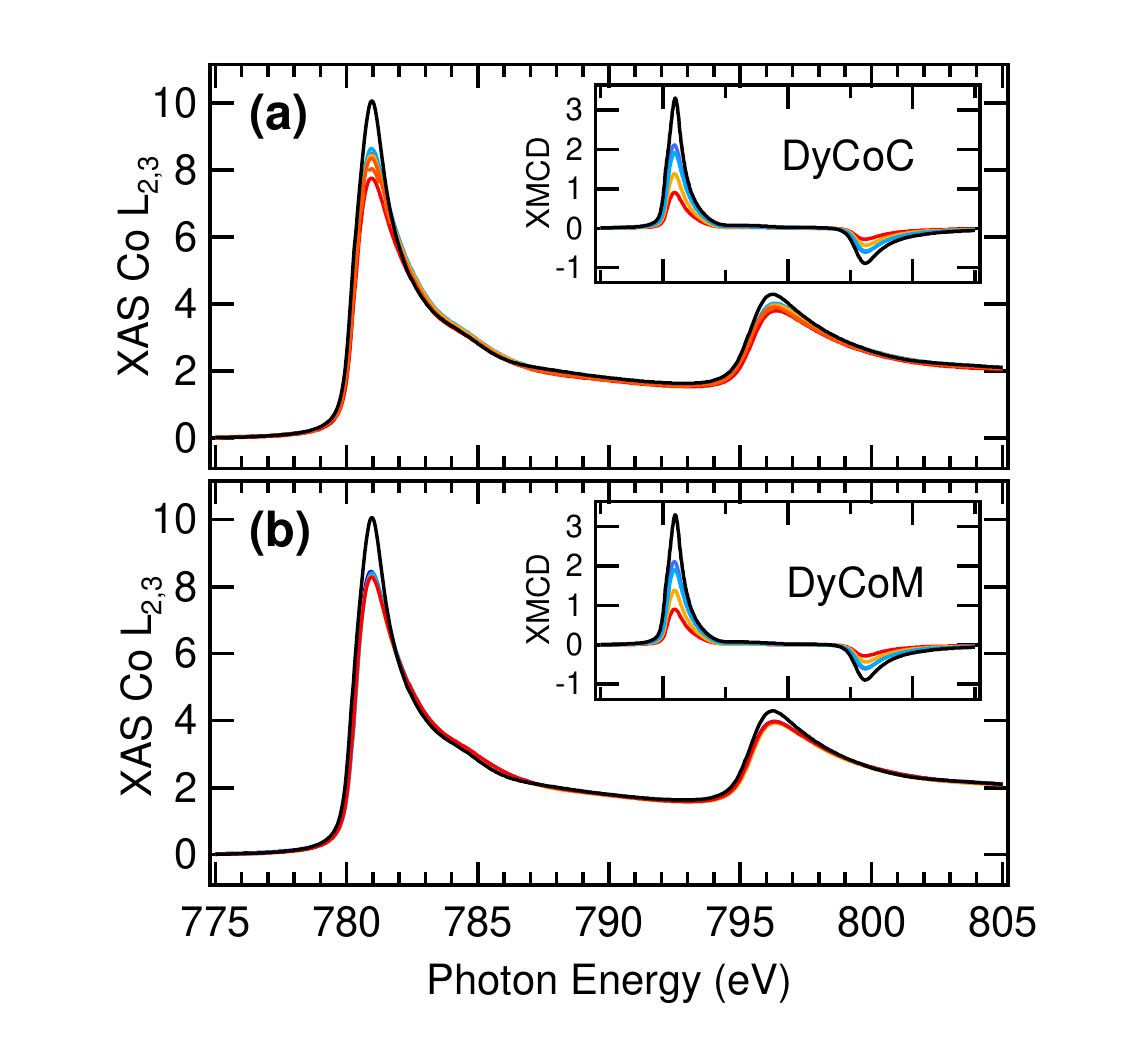}
\caption{X ray absorption (XAS) spectra obtained at the Co $L_{2,3}$ edge of a pure cobalt thin film compared with the spectra of cobalt in (a) DCC  and (b) DCM thin films taken at 2K and 0$^{\circ}$ (blue), 20$^{\circ}$ (green),45$^{\circ}$ (orange), and 70$^{\circ}$ (red) beam incidence angles (magnetic field orientation angle) with respect to the normal to the sample. The most intense spectrum (black line) is from pure cobalt. In the inset, the XMCD spectra taken at different orientation field angles, using the same code color.\label{fig_XAS_DCC_Co_comp}}
\end{figure}

The magnetic moments of cobalt were deduced from their $L_{2,3}$ spectra. The method used to extract the cobalt absorption coefficient to correctly apply the XMCD sum rules \cite{XMCD_L_SumRule_Altarelli,Carra_PRL}, which considers saturation effects, is fully described in \cite{NdCo_PRB}. The number of holes was calculated comparing their unpolarized absorption spectra with that of a pure cobalt reference sample deposited in similar conditions than the rest of the films. The number of holes for this reference sample was set to the tabulated for pure cobalt, 2.49, yielding a magnetic moment at 2 K (1.79$\pm$0.02$\mu_{B}$) similar to the measured by others (1.77 $\mu_{B}$) \cite{Co_XMCD_ChenPRL}. As it can be observed from figures \ref{fig_XAS_DCC_Co_comp}(a) and \ref{fig_XAS_DCC_Co_comp}(b), the shape of the spectra was almost identical to the cobalt reference except in their intensity which was lower and almost independent of the sample orientation angle or temperature. Their number of holes reduced to approximately 91$\%$ those of pure cobalt. The total magnetic moment of the samples together with the related orbital and spin components and orbital to spin ratios are shown in table \ref{table_Co_moments}.  The total magnetic moment of the two alloys at 2 K and normal incidence (easy magnetic axis) was similar within the error, of the order of 1.33$\pm 0.03 \mu_{B}$. This value is in agreement with the expected from those observed in NdCo alloys as a function of the number of holes \cite{NdCo_PRB}, and they would correspond to a RE concentration higher than the nominal, of the order of DyCo$_{3.5}$. The magnetic moment obtained at RT (only measured at normal incidence in DCC) was practically the same than at 2 K, 1.34$\pm 0.03 \mu_{B}$.  

\begin{table}[h]
\caption{Total magnetic moment (m$_{tot}$), efective spin (m$^{*}_{s}$) moment, orbital (m$_{o}$) moment, and its ratio m$_{o}$/m$^{*}_{s}$, of cobalt obtained by XMCD in DyCo alloys (samples DCC and DCM) and YCo alloys (YCC and YCM). All moments are given in $\mu_{B}$ units. m$^{*}_{s}$ and therefore, m$_{tot}$ have not been corrected by the m$_{Tz}$ term, the dipole moment of spin. Error bars are of the order of 2$\%$ for $m_{s}$ and $m_{o}$.
As a reference, pure cobalt (thin film) at 2 K: m$_{s}=$1.58 $\mu_{B}$; m$_{o}=$0.21 $\mu_{B}$; m$_{tot}=$1.79 $\mu_{B}$\label{table_Co_moments}}
\begin{ruledtabular}
\begin{tabular}{lllllll}
Sample & T & & \multicolumn{4}{l}{Field Orientation Angle } \\	%\cline{7,7}
\colrule\hline
&&&0$^{\circ}$&20$^{\circ}$&45$^{\circ}$&70$^{\circ}$ \\ \hline
DCC& RT &m$^{*}_{s}$&1.22  & - & - &1.11\\  
& &m$_{o}$& 0.12 & -& - &0.16 \\
& &m$_{tot}$& 1.34 & - &-&1.25 \\
& &m$_{o}/$m$^{*}_{s}$& 0.09 &  -& -& 0.12\\	
& 2 K &m$^{*}_{s}$& 1.17 & 1.15 & 0.91&0.41 \\
& &m$_{o}$& 0.16 & 0.10 &0.08  &0.06 \\
& &m$_{tot}$& 1.33 & 1.25 &0.99 &0.50 \\
& &m$_{o}/$m$^{*}_{s}$& 0.14 & 0.09 &0.09 &0.10 \\	\hline
DCM & RT &m$^{*}_{s}$&- & - &- & 1.08\\
& &m$_{o}$& - & - & - &0.16\\
& &m$_{tot}$& - & - & -& 1.24\\
& &m$_{o}/$m$^{*}_{s}$& - & - & - & 0.15\\
& 2 K &m$^{*}_{s}$&1.15  & 1.10 &0.79 &0.54 \\
& &m$_{o}$&0.15  & 0.10 &0.06  &0.04 \\
& &m$_{tot}$&1.30  &1.20  &0.85 &0.58 \\
& &m$_{o}/$m$^{*}_{s}$&0.13  & 0.09 & 0.08 &0.09 \\	\hline
YCC & 2 K &m$^{*}_{s}$&1.28  & - & 1.24& 1.28\\
& &m$_{o}$&0.14  & - & 0.19&0.16 \\
& &m$_{tot}$&1.42  & - & 1.43&1.44 \\
& &m$_{o}/$m$^{*}_{s}$&0.11  &- & 0.16 &0.13 \\	\hline
YCM & 2 K &m$^{*}_{s}$& 1.26 & - &1.18 &1.22 \\
& &m$_{o}$& 0.07 & - &0.18 &0.17 \\
& &m$_{tot}$&1.33  & - &1.36 &1.39 \\
& &m$_{o}/$m$^{*}_{s}$&0.06  & - &0.15 &0.14 \\

\end{tabular}
\end{ruledtabular}
\end{table}

The effective spin and orbital magnetic moments of cobalt changed with the field orientation angle at RT. The effective spin moment m$_{eff}$ obtained from the sum rules is the sum of the spin moment $m_{s}$ and the dipolar moment $m_{T}$ which is angle dependent:

\begin{equation}
m^{\theta}_{eff}=m^{\theta}_{s}-7m^{\theta}_{T}\label{eq_eff_spin_moment}
\end{equation} 

The dipolar moments at normal and in plane field orientation are -0.013 $\mu_{B}$ and 0.004 $\mu_{B}$, respectively, assuming a zero value for the dipolar moment when it is angle averaged\cite{Stohr_XMCDAngle,Stohr_JM3review}. The orbital anisotropy is also oriented in the plane, $\Delta$m$_{L}=$m$_{0^{\circ}}$-m$_{70^{\circ}}=$-0.02. The anisotropy of this term is usually linked to the anisotropy in the orbital moment \cite{Stohr_JM3review,vanderlaan,theory_japos2019} as observed in other 5d-3d systems like Au/Co/Au \cite{Stohr_CoAu}. This means that the cobalt sublattice of the alloys have an in plane magnetic anisotropy at RT, in coincidence with what we observed in NdCo alloys \cite{NdCo_PRB}. Actually, the in plane anisotropy of the cobalt sublattice is the expected for the oblate shape of the Nd and Dy $4f$ orbital, which is the one that generates the expected crystal field orientation for the RE out of plane anisotropy. Nevertheless, the anisotropy of the observed values is small, with values below the experimental error. This would discard cobalt as the origin of the PMA of the alloy at RT. 

\begin{figure}
\includegraphics[width=8 cm]{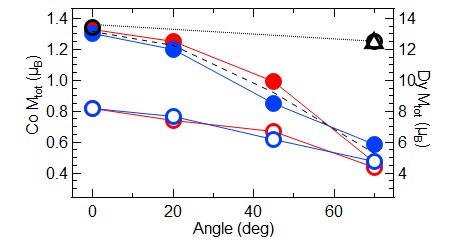}
\caption{ Total magnetic moment of cobalt (solid dots) and dysprosium (empty dots) in samples DCC (red dots) and DCM (blue dots) measured at 2 K. The black dashed line in (a) is the fit to a cosine function plus a constant. Black dots an triangles are cobalt moments measured at RT. The estimated error of the measurement is the width of the dots. \label{Fig_Co_Dy_CodDy_vs_angle}}
\end{figure}

The total magnetic moment measured at 2 K is displayed in figure \ref{Fig_Co_Dy_CodDy_vs_angle}(a) as a function of the field orientation angle. It decreases in both samples with increasing field angle orientations because of the uncompleted magnetic saturation of the alloy at the 6 T applied field. Due to the higher strength of the exchange coupling between cobalt atoms compared to the indirect exchange interaction with the dysprosium moments, the orientation angle dispersion of their magnetic moments is expected to be much smaller than those of the dysprosium atoms. The dashed curve displayed in figure \ref{Fig_Co_Dy_CodDy_vs_angle}(a) is the fit of the averaged magnetic moments of the two samples using a cosine function plus a constant. The constant could be interpreted as the portion of the cobalt magnetic moment that it is oriented parallel to the field. This constant is only 10 $\%$ of the total intensity, indicating that most of the cobalt magnetic moments were fixed at the easy axis direction. 

The orbital magnetic moment at 0$^{\circ}$ field orientation was bigger than at RT, increased from 0.12 $\mu_{B}$ to 0.15 $\mu_{B}$. Both samples showed similar values. For the rest of the field orientation angles, only the orbital to effective spin moments ratio can be compared because the uncompleted magnetic moment saturation. This ratio is the highest at  0$^{\circ}$ and it decreases significantly in both samples, from 0.14 at normal orientation, to 0.09 (0.10 at 70 $^{\circ}$ in sample DCC). This indicates that the perpendicular magnetic anisotropy of the alloy involves either an anisotropy in the orbital moment of cobalt in the same direction of the easy axis or a larger spin component at angles far from the easy axis. 

In order to check that if this effect is caused by the magnetic interaction at cobalt with the RE, YCo films deposited in similar conditions of concentration, film thickness and deposition process were measured by XMCD. The chemical interaction of Yttrium is equivalent to that of RE, with a nearly empty $4d$ band, but without an occupied $4f$ orbital, i.e., with no $4f$ magnetic moment. Actually, their RE-Co phase diagram is very similar \cite{YCo_PhaseDiagram,DyCo_PhaseDiagram}. Therefore, the atomic structure of the YCo alloy should be similar to that of DyCo. The magnetic moments of cobalt in YCo measured at RT (normal orientation only) and 2 K are displayed in table \ref{table_Co_moments}. The values are about 10 $\%$ smaller in the multilayer than in the continuous film and they increase in both films when temperature goes from RT to 2 K, in contrast to the observed in the DyCo alloys where there was almost no difference. The YCo alloys were magnetically soft and their magnetic anisotropy was in the plane. Their orbital to effective spin ratios were similar to the found at normal orientation in the DyCo films. But they did not show the strong decrease in their value with the increasing orientation angle as the observed in DyCo. This fact, together with the cosine like variation in the total magnetic moment of cobalt in the DyCo films, indicates that their probed cobalt atoms are mostly located at magnetically anisotropic environments. By comparison with the YCo alloy, the magnetic anisotropy found in the cobalt sublattice of the DyCo films must be induced by the dysprosium atoms. It would be expected, therefore, a similar angle variation in the magnetic moment of the dysprosium sublattice as the found in cobalt.

\subsection{XMCD Dy \label{XMCD Dy}}
\begin{figure}
\includegraphics[width=8 cm]{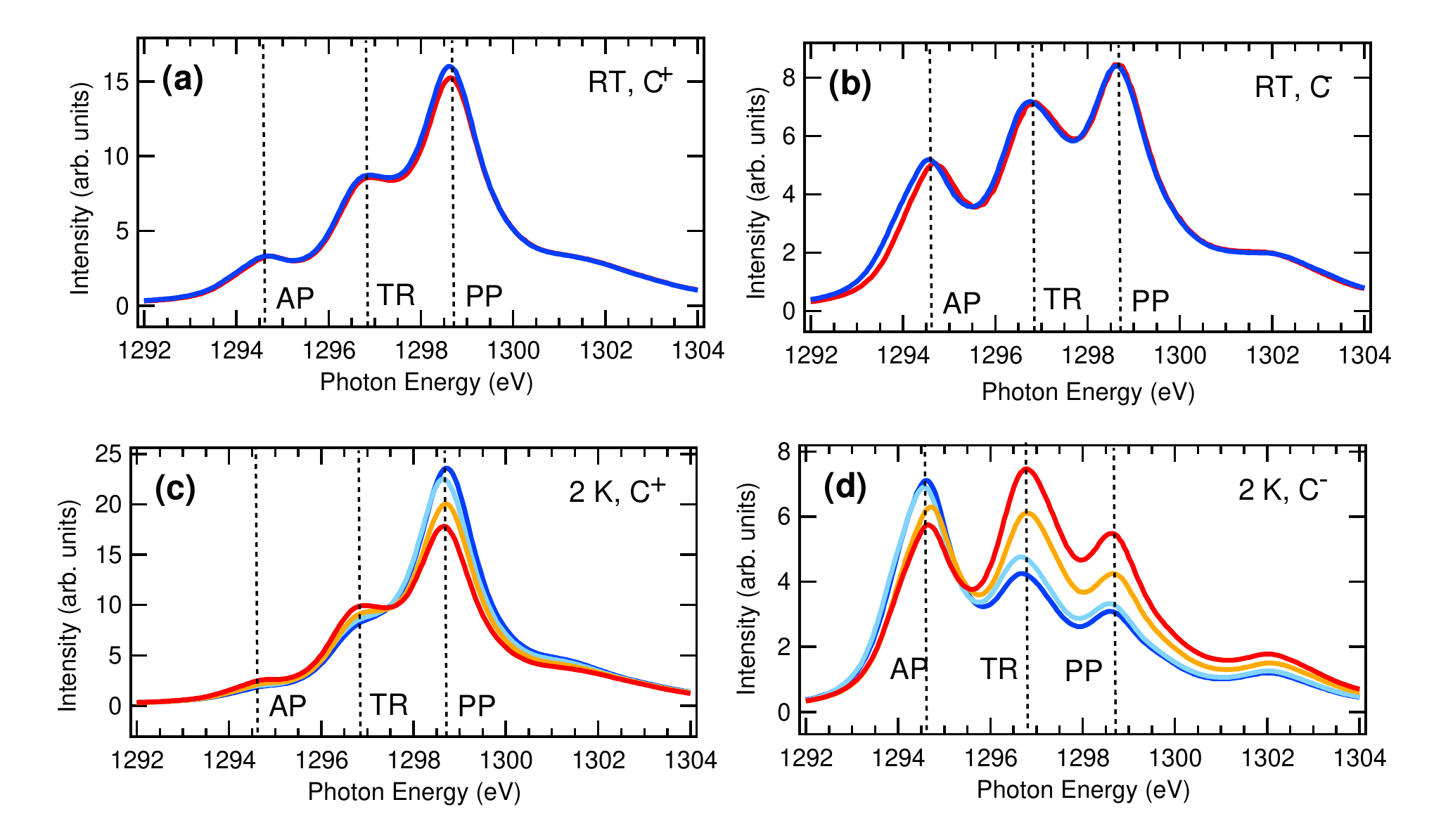}
\caption{ Dy $M_{5}$ spectra of sample DCM obtained at RT (spectra at (a) and (b)) and  2 K (spectra (c) and (d)). The spectra taken with right handed circular polarization are (a) and (c). (b) and (d) are the spectra taken using left handed circular polarization. RT spectra were taken at 0$^{\circ}$ (dark blue line) and 70$^{\circ}$ (red line) incident angles. 2 K spectra were taken at 0$^{\circ}$ (dark blue line), 20$^{\circ}$ (light blue line), 45$^{\circ}$ (orange line) and 70$^{\circ}$ (red line) incident angles.    \label{Fig_XAS_Cmyp_RT_2K}}
\end{figure}

In dysprosium, the most relevant changes in the spectra occur in the Dy $M_{5}$ edge. Figure \ref{Fig_XAS_Cmyp_RT_2K} shows the Dy $M_{5}$ taken in sample DCM at RT and 2 K, for the two circular polarizations, at different angles. Sample DCC showed similar changes. The Dy $M_{5}$ spectral shape is defined by three intense peaks, marked in the figure as peaks AP, TR and PP, which are located at photon energies 1294.5 eV, 1296.7 eV, and 1298.6 eV, respectively. Their relative intensity changes with the polarization sign and with the field orientation angle. At RT, the changes with the angle are small, but they are very evident at 2 K. At this temperature, peak PP is the most intense at $C^{+}$ polarization and normal incidence. Within the same polarization, peak PP decreases in intensity as the field orientation angle (incident X ray beam angle) increases. The opposite is observed using $C^{-}$ polarization. In this case, peak marked AP is the most intense at normal incidence. For both polarizations, peak TR increases in intensity when the incident angle increases. As it will be shown, peaks PP, AP and TR are associated to electronic transitions to states where the magnetic moment of the related $4f$ orbital is parallel, antiparallel or transverse to the circularly polarized x ray beam wave vector, respectively. It is important to notice that, at RT, peak PP is the most intense also in $C^{-}$ polarization. This is opposite to what happen at 2 K, specially at 70$^{\circ}$, where the measured magnetic moment of dysprosium is the lowest and comparable to the measured at RT.

The XMCD spectra of Dy $M_{4,5}$ edge were analyzed following the process explained in \cite{NdCo_PRB}. Table \ref{table_XMCD_DyCo} shows the magnetic moments of dysprosium in each of the samples at RT and 2 K. The relative weakness of the AF interaction with cobalt and the structural disorder of the alloy makes the magnetic moment of the dysprosium atoms to be dispersed in its orientation with respect to the cobalt magnetic moment forming what is called an asperomagnet \cite{amorphous_magnets_Coey}. The moment orientation of the RE atoms when the alloy is magnetized in the easy axis is usually taken as uniform and symmetrically distributed around this axis within a cone with a half opening angle $\theta_{C}$ ($\theta_{C}=\pi/2$ represents the distribution of moments in a semisphere). For field orientations away from the easy axis, a dispersion in the shape and opening angle of the cone is expected. In this case, we assume that the resulted distribution of the dysprosium can be approximated by a cone that has its symmetry axis deviated from the field direction an angle $\varphi_{C}$. The magnetic moment measured is the averaged sum within the cone of the magnetic moment component along the field orientation axis. At the easy axis, this sum is $M_{Dy}\cos^{2}\theta_{C}/2$, where $M_{Dy}$ is 9.8 $\mu_{B}$, the dysprosium magnetic moment of its $4f$ orbital. The magnetic anisotropy would then change the value of the measured magnetic moment by opening the cone angle and changing its symmetry axis orientation. 

\begin{table}[h]
\caption{Efective spin moment,m$^{*}_{s}$, orbital moment, m$_{o}$, dipole moment of spin, m$_{Tz}$, and total magnetic moment, m$_{tot}$, of dysprosium obtained by XMCD in samples DCC and DCM. All moments are given in $\mu_{B}$ units. The estimated error bars is of 2$\%$ in m$^{*}_{s}$, m$_{o}$ and m$_{Tz}$. m$_{Tz}$=$\frac{m_{o}}{6}\left[\frac{m^{*}_{s}}{m_{o}}-\frac{m_{s}}{m_{o}}\right]$ \cite{XMCD_US}, using the theoretical value of $\frac{m_{s}}{m_{o}}=0.475$ \cite{teramura_RE}. m$_{tot}$=m$_{o}$+(m$^{*}_{s}$-6m$_{Tz}$)\label{table_XMCD_DyCo}}
\begin{ruledtabular}
\begin{tabular}{lllllll}
Sample & T & & \multicolumn{4}{l}{Field Orientation Angle } \\	%\cline{7,7}
\colrule\hline
&&&0$^{\circ}$&20$^{\circ}$&45$^{\circ}$&70$^{\circ}$ \\ \hline
DCC & RT &m$^{*}_{s}$&-1.92 & - & - &-1.92\\
& &m$_{o}$& -1.42 & -& - &-1.41 \\
& &m$_{Tz}$& -0.09 &  -& -& -0.10\\	
& &m$_{tot}$& -2.78 & - &-&-2.74 \\
& 2 K &m$^{*}_{s}$& 2.78 & 2.51 & 2.19& 1.46 \\
& &m$_{o}$& 4.19 & 3.81 &3.42 &2.23 \\
& &m$_{Tz}$& 0.26 & 0.23 &0.19 &0.13\\
& &m$_{tot}$& 8.18 & 7.43 &6.67&4.35 \\ \hline
DCM & RT &m$^{*}_{s}$&- & - &- & -0.91\\
& &m$_{o}$& - & - & - &-1.44\\
& &m$_{Tz}$& - & - & - &-0.07\\
& &m$_{tot}$& - & - & -& -2.81\\
& 2 K &m$^{*}_{s}$&2.84 &2.64 &2.14 &1.61 \\
& &m$_{o}$&4.21  & 3.93 & 3.17  &1.61 \\
& &m$_{Tz}$&0.28  & 0.26 & 0.21 &0.15 \\	
& &m$_{tot}$&8.2  &7.66  &6.19 &4.73 \\

\end{tabular}
\end{ruledtabular}
\end{table}

At RT, the dysprosium magnetic moment has almost no difference at both orientations, as expected from the hysteresis loops. The PMA of dysprosium has not enough energy to distort its magnetic moment distribution cone at RT. There is not a significant variation between the values obtained in both samples, which are about -2.8 $\mu_{B}$ (sample DCM only measured at 70$^{\circ}$ field orientation). This quantity is small. The lowest magnetic moment that can be measured assuming that all dysprosium atoms are AF coupled to Co, i.e., uniformly distributed within a cone of half opening angle $\theta_{C}=\pi/2$ (semisphere), is $M_{Dy}/2=$ -4.9 $\mu_{B}$, i.e., half the value of its total magnetic moment. This means that there should be dysprosium atoms with their moment orientation opposite to that of those AF coupled to cobalt. Theoretical calculations estimated that the averaged magnetic moment of Dy decays to half its value at about 300 K in crystalline DyCo$_{5}$ \cite{2018_Tc_RETM}. The lower values observed in the present case should have to do with the intrinsic disorder of the measured alloys, with dysprosium atoms distributed in different atomic environments where, in some of them, the exchange coupling to cobalt should be weaker than in crystalline DyCo$_{5}$ or possibly nonexistent. The analysis of the Dy $M_{5}$ spectra exposed in a later section searches to understand how these dysprosium atoms are distributed through the depth proved by TEY and how this can affect the TEY measurements including the possible presence of segregated dysprosium at the interface with the aluminum capping layer.

Figure \ref{Fig_Co_Dy_CodDy_vs_angle} (right side scale) displays the variation of the dysprosium magnetic moment as a function of the field orientation angle at 2 K compared to that of cobalt. Both samples have similar values which decay with the angle at a slower pace than in the cobalt sublattice. This magnetically less anisotropic behavior of dysprosium is somehow contrary to what expected: the magnetic moment of the RE should be, compared to cobalt, the one fixed with the highest energy to the easy axis since the PMA in these alloys must stems from it, as deduced from the analysis done in the previous section (\ref{Co spectra}). The total magnetic anisotropy of the alloy should depend on the distribution of the crystal field orientation at the RE sites\cite{SKOMSKI_AniRE}, and its influence on the magnetic moment of the bonding TMs at each RE site. Following the single ion anisotropy model \cite{SKOMSKI_AniRE}, both crystal field and RE-TM indirect exchange are expected to be tightly related. Therefore, the RE environments that most contribute to the magnetic anisotropy of the alloys should be strongly exchange coupled to the TM as well.

The magnetically less anisotropic character of the measured dysprosium sublattice indicates that only a portion of the dysprosium atoms probed by TEY should be active magnetic anisotropy generators. The magnetic field felt by dysprosium is the sum of the external applied field and the molecular field, which is mainly provided by the interatomic exchange interaction with cobalt atoms, whose strength can be estimated in, at least, more than 150 T \cite{NdCo_PRB}. As it has been observed, the rotation of the cobalt moment at grazing orientations is small. Therefore, the effective molecular field felt by the dysprosium should be closer to that of the applied field at those sample orientation angles, meaning that a portion of the probed dysprosium should be somehow decoupled or poorly coupled to cobalt, as stemmed from the analysis done at RT. One of the challenges in the analysis of the Dy $M_{5}$ spectra, shown in the next sections, is to determine the location of the dysprosium which are apparently weakly interacting to cobalt.

A way to compare the magnetization measured in the bulk with the magnetization measure near the surface (XMCD using TEY) is to evaluate how these magnetic moments found by XMCD fit with the deduced from VSM, shown in table \ref{tab_Comp_Temp_DCC_DCM}. When the magnetic moment of the cobalt used to calculate the total magnetization of the alloy is the deduced by XMCD, 1.33 $\mu_{B}$, the magnetic moment of dysprosium that it is needed to match the magnetization measured by VSM at RT is 3.8 $\mu_{B}$, 1 $\mu_{B}$ above the measured by XMCD. This value is still below the expected if all dysprosium magnetic moments were AF coupled to cobalt. When the same estimation is done at 2 K, the magnetic moment expected for dysprosium is 7 $\mu_{B}$, 1 $\mu_{B}$ smaller than the found by XMCD. These differences between bulk (VSM) and surface (XMCD) are similar to the reported in \cite{SciRep_2015,SciRep_2019}. These were justify there because it was assumed a higher dysprosium concentration at the region near surface than in the bulk due to RE segregation. If that was happening in our samples, their cobalt and dysprosium moments in the bulk should be higher than the measured by XMCD. For instance, if the cobalt moment in the bulk raised to 1.50 $\mu_{B}$, the related dysprosium moment should be 4.7 $\mu_{B}$, which is almost 2 $\mu_{B}$ higher than the measured by XMCD. This would set the surface at a higher $T_{Comp}$ than in the bulk. But none of our samples showed wing side hysteresis loops at 7 T at temperatures near the $T_{Comp}$, as the reported in \cite{SciRep_2015,SciRep_2019} in alloys of similar thickness and concentration. 

The $T_{Comp}$ of a ferrimagnetic RE-TM alloy essentially depends on the magnetic exchange strength between the RE and the TM atoms. If the segregated RE atoms at the region near the surface do not bond tight to cobalt, their exchange interaction will be weaken. The effect, in terms of magnetic moments, will be similar to an increase of the $T_{Comp}$ because, above this temperature, there will be dysprosium moments that will not be AF coupled to cobalt, reducing the total moment of the dysprosium sublattice. Below $T_{Comp}$, the effect will be the opposite, i.e., they will add up in moment to the moment of the AF coupled-to-cobalt dysprosium. This is not an unlikely situation regarding the magnetic moments of dysprosium measured by XMCD, which are well below the expected if all RE atoms were AF coupled to cobalt. Moreover, if RE segregation occurs is because it favors RE-RE metallic bonding. To demonstrate the presence of these two kinds of RE atoms requires a technique able to detect them. The next section shows a way to do it in a quantitative way by analyzing the circularly polarized Dy $M_{5}$ spectra. Actually, dysprosium is specially well adapted for this kind of analysis, although it should be applicable to most of the REs, with the exception of those with $J=0$ like $Gd^{+3}$.

%The shape and intensity of the circularly polarized Dy $M_{4,5}$ spectra depends on the orientation of the dysprosium magnetic moments with respect to the direction of the beam, which is always parallel to the field orientation. Therefore, these spectra should contain information about the different possible magnetizations of the dysprosium atoms in the alloy. The following section presents a method to untangle the parallel (PP), antiparallel (AP) and transverse (TR) spectral related components of $J_{z}$ from the Dy $M_{4,5}$ spectrum, which will help to understand the reason behind the apparent magnetic decoupling between the dysprosium and the cobalt sublattices.

\section{Spectral analysis Method for Dy $M_{4,5}$ spectra\label{Spectral analysis Method for Dy $M_{4,5}$ spectra}}
Like in most of the RE, the electrons in the $4f$ orbital of dysprosium are well screened by the valence band orbitals and behave as in an isolated atom. Actually, the Dy $M_{4,5}$ edge spectrum, which involves electronic transitions form the 3d to the $4f$ orbital, is practically insensitive to the dysprosium chemical environment and it is well fitted by calculating it using only intratomic interactions\cite{Thole_PRB, Goedkoop_thesis}. The relative intensity of these transitions can be notably influenced by the bonding with other magnetic atoms through its indirect exchange magnetic interaction and the resulting crystal field by breaking its degeneracy in the $J_{z}$ momentum. Their relative weakness permits to treat them as perturbations whose most important effect is to orient the magnetic moment of $4f$ orbital along a specific direction, i.e., to break the degeneracy of the quantum number $M$ of the $4f$ angular moment component $J_{z}$.

Therefore, to a good approximation, only the intensity, but not the shape, of the Dy $M_{4,5}$ lines are modulated by the orientation of the total angular moment of the $4f$ orbital with respect to the direction of the polarized X ray. The allowed transitions in the dipole approximation are those in which $\Delta J=0,\pm1$. If light is circularly polarized, only $\Delta m=\pm1$ transitions are allowed. Therefore, if the beam is perfectly oriented parallel (antiparallel) to the magnetic moment of the $4f$ orbital, the resulting $M_{4,5}$ spectrum will be built with only those excitations where $\Delta J=0,\pm1$ and $\Delta m=1 (-1)$. We will call to this spectral components PP if $\Delta m=1$ and AP when $\Delta m=-1$. The circular dichroism spectrum, XMCD, is the difference between these two spectrum. The sum rules related to these transitions gives rise to the XMCD sum rules that permits to determine both the magnetic spin and orbital moment of the $4f$ orbital. $\Delta m=0$ transitions occurs when the light is linearly polarized along the magnetic moment direction. This spectral component is not present in the XMCD spectrum because depends on $\left\langle M^{2}\right\rangle$ \cite{Goedkoop_thesis}. We will call to this component TR.

The excitations of all the three spectral components PP, AP and TR occurs when the $4f$ magnetic moment and the incident circularly polarized light are not parallel. This is because the electric field felt by the $4f$ electrons is equivalent to the sum of two circularly polarized fields with opposite helicities, and a linear polarized field aligned with the direction of the magnetic moment. The detailed calculation of the amplitude of these fields is explained in appendix \ref{ApA}. The dependency of the amplitude of these polarized field with the magnetic moment orientation angle $\theta$ is:

\begin{equation}
P=\frac{e^{\pm i\varphi}}{\sqrt{2}}\left[\frac{(\cos \theta \pm 1)}{\sqrt{2}}P^{-1}_{z_{m}}+\frac{(\cos \theta \mp 1)}{\sqrt{2}}P^{1}_{z_{m}}+P^{0}_{z_{m}}\sin \theta\right]
\label{eq_dip1}
\end{equation}

\begin{figure}
\includegraphics[width=8 cm]{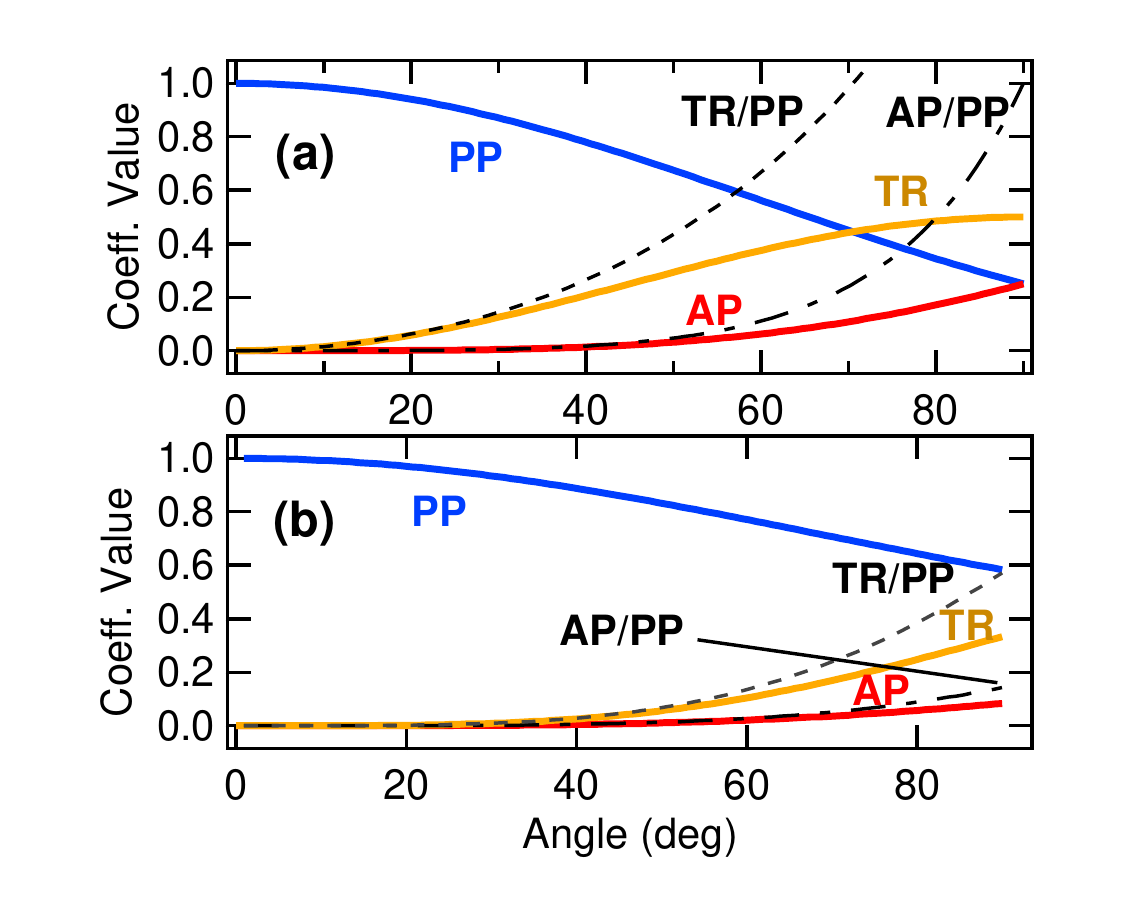}
\caption{ Angle coefficients for PP, AP and TR as a function of (a) magnetic moment orientation angle with respect to the circularly polarized incident beam, $\theta$, and (b) cone half opening angle $\theta_{C}$. \label{Fig_coeff_ang_conang}}
\end{figure}

$P^{-1}_{z_{m}}$, $P^{1}_{z_{m}}$ and $P^{0}_{z_{m}}$ are the dipole operators for right and left circular polarizations, and linear polarization along the $z$ axis, respectively. Taking into account the previous expressions, the modulation of the intensity of each of the spectral components (PP, AP and TR) with the orientation angle $\theta$ is given by the following expression (see appendix \ref{ApA}):

\begin{widetext}
\begin{equation}
PP+AP+TR=\left[\frac{(\cos\theta \pm 1)^{2}}{4}A^{-1}_{JJ'}+\frac{(\cos\theta \mp 1)^{2}}{4}A^{1}_{JJ'}+\frac{\sin^{2}\theta}{2}A^{0}_{JJ'}\right]
\label{eq_dip2}
\end{equation}
\end{widetext}

where $A^{q}_{JJ'}$ are the reduced angular integers. They are directly related to the transitions $\Delta J=0,\pm1$ and $q=\Delta M$. The distribution in energy and intensity of these transitions are well tabulated, both theoretically and experimentally \cite{Goedkoop_thesis,Goedkoop_PRB, Thole_PRB}.
Figure \ref{Fig_coeff_ang_conang}(a) displays the coefficients that multiply to the angular integers $A^{q}_{JJ'}$ showed in \ref{eq_dip2} as a function of the angle $\theta$. As expected, the coefficients that multiply to the TR ($A^{0}_{JJ'}$) and AP ($A^{1}_{JJ'}$) components increase their value only at angles closer to ${\pi/2}$.  

The most common situation found in the studied alloys, which have no defined crystal structure, is the one in which the orientation of the magnetic moment of dysprosium is not well defined but it is distributed over a range of angles due to structural and thermal disorder. This distribution is assumed to be uniform and symmetrical with respect to the axis of a cone with a half opening angle $\theta_{C}$. This distribution modifies the coefficients of the angular integers. The new coefficients in this situation are obtained by averaging the angle coefficients in \ref{eq_dip2} for each component within this angle, resulting in the following expression as a function of $\theta_{C}$:

\begin{widetext}
\begin{equation}
PP+AP+TR=\frac{1}{3}\frac{\left(1-\cos^6 \frac{\theta_{C}}{2}\right)}{\sin^2 \frac{\theta_{C}}{2}}A^{-1}_{JJ'}+\frac{1}{3}\sin^4 \frac{\theta_{C}}{2}A^{1}_{JJ'}+\frac{1}{3}\frac{\left(\cos^3 \theta_{C}-3\cos \theta_{C}+2\right)}{1-\cos \theta_{C}}A^{0}_{JJ'}
\label{Int_Pol_angle}   
\end{equation}
\end{widetext}

Figure \ref{Fig_coeff_ang_conang}(b) shows the coefficients of each excitation component as a function of the cone half opening angle $\theta_{C}$. The coefficients of the TR and AP components increment its value at large angles, but not as fast as in the case of a defined angle orientation. This is better observed when the ratio between the TR and AP components, shown in both figures, are compared. These ratios are substantially smaller in the cone angle coefficients (figure \ref{Fig_coeff_ang_conang}(b)).  

When the cone has an inclination angle $\varphi_{C}$ with respect to the beam direction, the components PP' AP' and TR' of the magnetic moment in the cone are projected in the beam axis following expression \ref{eq_dip2}, substituting the angle $\theta$ by $\varphi_{C}$. The PP' component is calculated using equation \ref{Int_Pol_angle}, the AP' component is the same as the equation \ref{eq_dip2}, but interchanging the factors $A^{-1}_{JJ'}$ by $A^{1}_{JJ'}$, and the TR' component is calculated by multiplying TR by $\sin^{2}\theta_{C}/2$.

Since the cross section values of the angular integers are tabulated \cite{Thole_PRB,Goedkoop_thesis} and the angle dependent modulation of the PP, AP and TR components are known (equations \ref{eq_dip2} and \ref{Int_Pol_angle})), it is possible to determine the orientation of the magnetic moment of any RE in any chemical environment just by identifying their PP, AP and TR components and checking their relative intensities. This is something that can be inferred also by comparing the magnetic moment deduced from the application of the XMCD sum rules and compare it with the total magnetic moment of the RE. But this is only valid if there is a single magnetic form of dysprosium. Things become more complicated if the analyzed material contains RE in different magnetic states, i.e., RE with different magnetic moment orientation distributions. The deconvolution of the spectra in their PP, AP and TR components is then required. This could be done by their direct calculation using numerical methods as the employed by Thole et al. \cite{Thole_PRB}. This is the approach used in \cite{deconvoluted_DyCo}. But it is important to notice that the build of the RE $M_{4,5}$ spectrum only requires, to a good approximation, the shape of the spectral component, which can be isolated by spectral methods. The Dy $M_{5}$ spectrum is specially suited for this kind of analysis.

\begin{figure}
\includegraphics[width=8 cm]{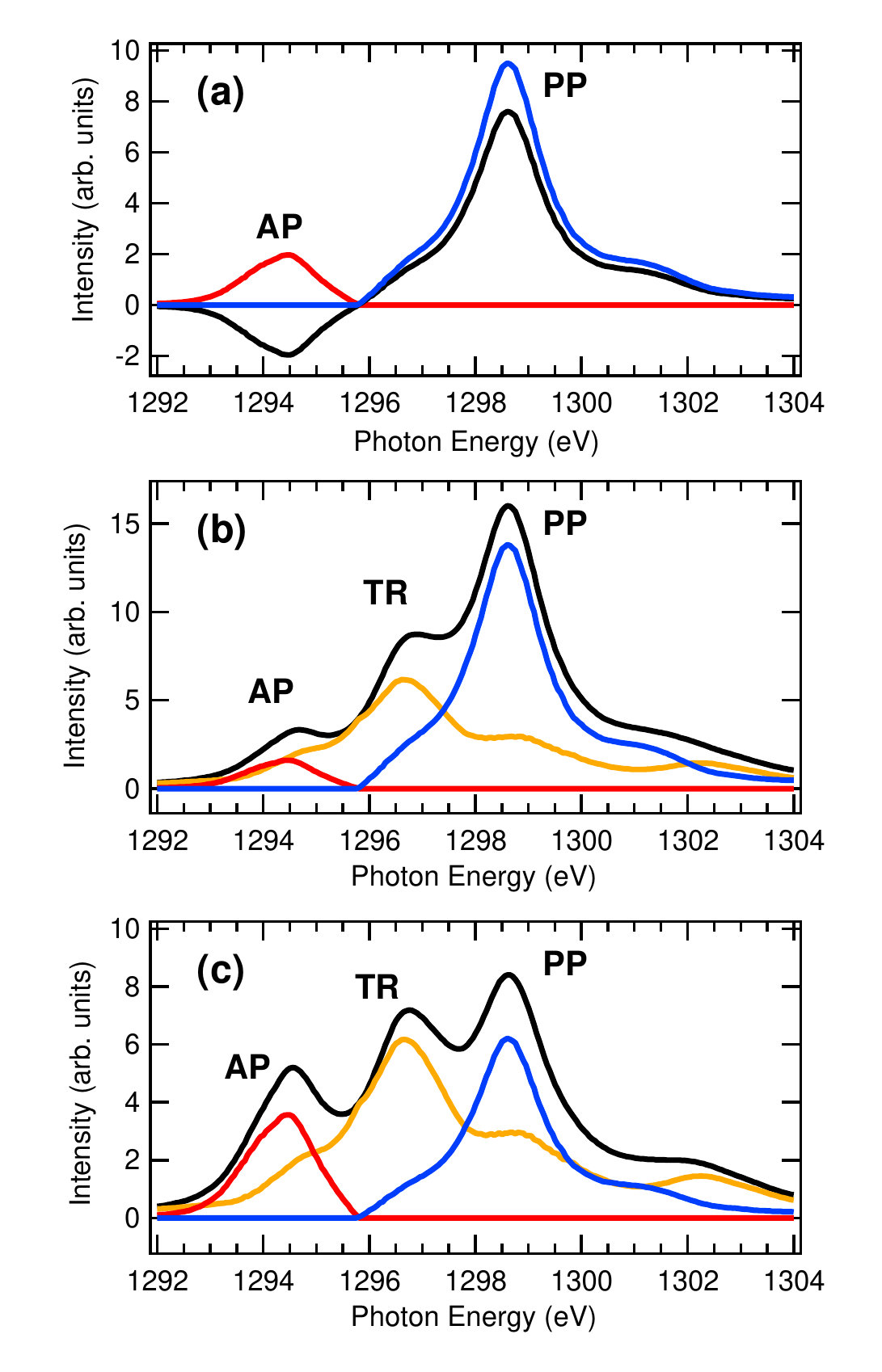}
\caption{ (a) XMCD spectrum of the DCC sample at RT  and normal incidence, and its related PP and AP components; (b) $C^{-}$ polarized spectrum of DCC sample taken at RT and normal incidence with their related PP, AP and TR components; (c) $C^{+}$ polarized spectrum of DCC sample taken at RT and normal incidence with their related PP, AP and TR components \label{Fig_XMCD_DCC_PP_AP_TR_RT}}
\end{figure}

Due to the large value of the orbital moment, $L$, of the $4f$ orbital in dysprosium, the splitting between the PP and AP components in the $M_{5}$ edge is the largest among the RE. Dy is the RE that has the lowest overlap between these two sets of excitations, which is estimated in less than $5\%$ \cite{Goedkoop_thesis}. This eases the extraction of both spectral components from any XMCD spectra at the Dy $M_{5}$ edge. Figure \ref{Fig_XMCD_DCC_PP_AP_TR_RT} shows the XMCD spectrum for Dy $M_{5}$. Only the PP and AP components are in the spectra, which are well distinguished because they have opposite sign. These two spectral shapes can be taken, as a first approximation, to the PP and AP true components of the spectra. The TR spectral component is extracted by subtraction of these two approximated components to the related circularly polarized spectra. For the subtraction, each of the components must be multiplied by a coefficient which is related to its relative spectral intensity, which depends on the sum of all the RE magnetic moment orientation distributions probed. These coefficients must be the same in both circularly polarized spectra, bearing in mind that the PP and AP components are interchanged in the two circular polarization spectra taken at the same magnetic field direction. The determination of the value of these coefficients is not precise, since it depends on how well the shape of the TR component is known. In our approach, we used the spectral lineshape theoretically calculated by Thole et al. \cite{Thole_PRB}. This so extracted TR component contains the overlap region between the PP and AP components. This produces some shape differences between the TR obtained at different beam orientations angles, because the proportion of the TR component respect to the PP and AP components changes. In our case,  the spectra taken at 70$^{\circ}$ has, clearly, the lowest overlap proportion in its TR component. Then, the PP-AP overlapping region can be extracted by subtracting this TR component to the TR component withdrawn at 0$^{\circ}$, being the former conveniently normalized to the intensity of the 0$^{\circ}$ TR component. This overlapping component is added to the PP and AP components directly extracted from the XMCD spectra obtained at any angle and temperature. The process is then repeated for the extraction of the related TR components, reducing in this way its overlapping portion, and approaching to the true TR component. 

\begin{figure}
\includegraphics[width=8 cm]{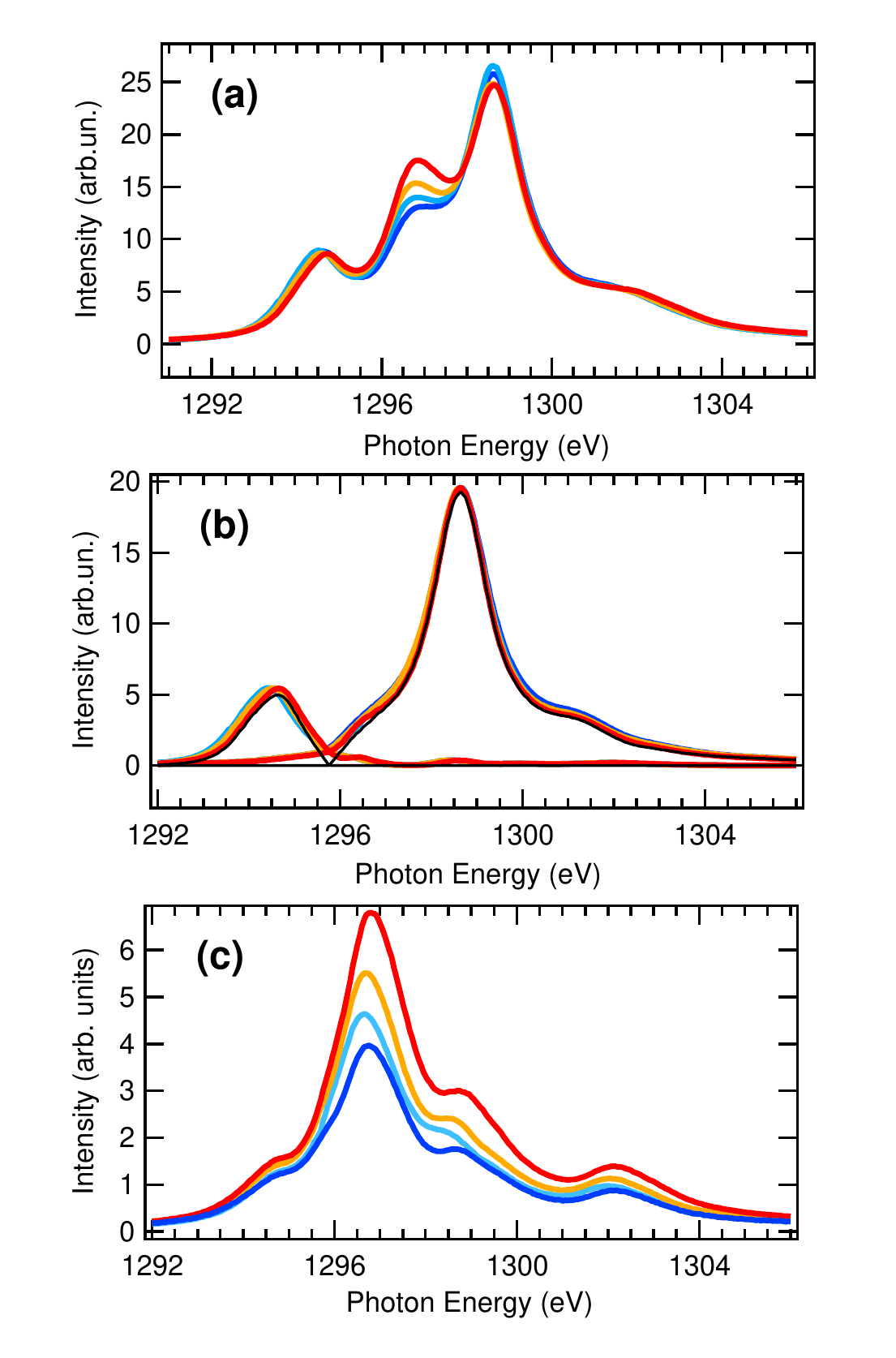}
\caption{ (a) XAS spectrum ($C^{+}+C^{-}$) of sample DCC taken at 2 K and at different orientation angles: 0$^{\circ}$ (dark blue line), 20$^{\circ}$ (light blue line), 45$^{\circ}$ (orange line) and 70$^{\circ}$ (red line). (b) PP and AP components, and (c) TR components extracted from the spectra of the DCC sample at 2 K. The colors code is the same as in (a). \label{Fig_XAS_PP_AP_TR_corrected}}
\end{figure}

The result of applying this process in our samples is shown in figure \ref{Fig_XAS_PP_AP_TR_corrected}. Although the overall shape of each of the components is very similar for any spectra, they were not exactly the same. For instance, we observed a reduction in the energy distance between the AP and PP peaks position with increasing field orientation angle, which is visible in their XAS spectrum shown in figure \ref{Fig_XAS_PP_AP_TR_corrected} (a) for sample DCC. The same observation was made in sample DCM and at RT. The reason of this is beyond the scope of this work. More specific experiments are required to determine if these differences, which are small in any case, are caused by experimental factors or because the spectra are actually sensitive to the chemical and/or crystal field environments \cite{EuO_PRL,NdCo_PRB,SmCo_anisotropy}. However, such differences imposed that the deconvolution of each spectrum and, later on, its model fitting, had to be done using its own PP, AP and TR components. We noticed that the obtained coefficients for each of the components used in the deconvolution of the spectra have little variations when using the corrected or the uncorrected spectral components. 

%To understand the errors introduced in the previously explained approach, the spectral fittings were done using spectral PP and AP components with and without the overlapping component, using the related TR component in each case. We will only 

\begin{figure}
\includegraphics[width=8 cm]{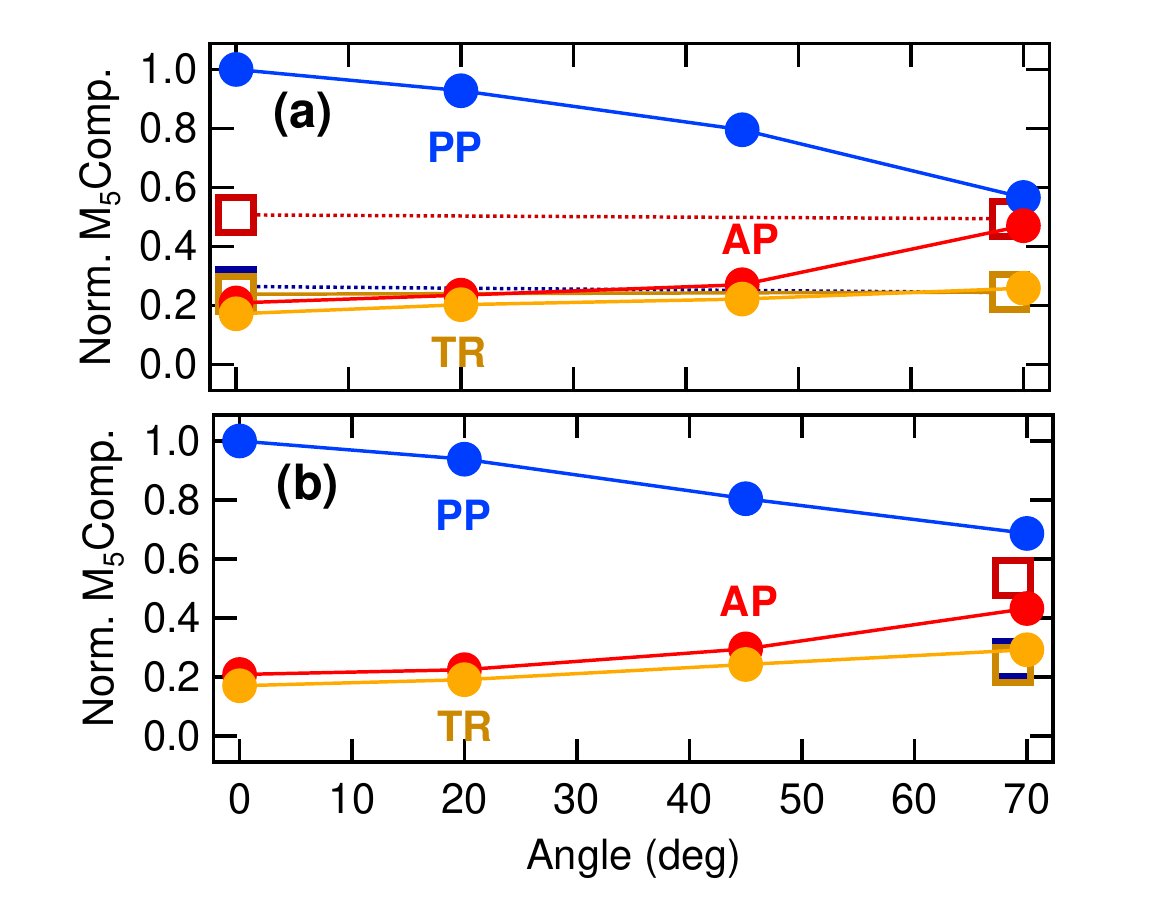}
\caption{ Normalized PP (blue dots), AP (red dots) and TR (yellow dots) components in Dy $M_{5}$ spectra of samples (a) DCC and (b) DCM taken at 2 K as a function of  the circularly polarized incident beam. Empty squares are related to the components of the Dy $M_{5}$ spectra taken at RT. \label{DCC_DCM_PP_AP_TR_vs_angle}}
\end{figure}

Figure \ref{DCC_DCM_PP_AP_TR_vs_angle} shows the normalized coefficients of the three spectral components obtained from the deconvolution of the spectra of the two samples DCC and DCM at RT and 2 K, as a function of the beam and magnetic field orientation angle. The plot was done to compare it with the angle dependent coefficients plotted in figures \ref{Fig_coeff_ang_conang}(a) and \ref{Fig_coeff_ang_conang}(b), which are cross section independent. The normalization of the coefficients was done by dividing all the components by the intensity of the PP component measured at normal incidence, which is the one with the highest intensity, and corrected by their relative cross sections. Cross sections were obtained from \cite{Goedkoop_thesis}. The used cross section ratio between the PP and the AP components were 3.4, and 1.85 between PP and TR components. The TR component is divided by 2, since there are two possible traversal directions to $J_{z}$. The coefficients obtained at 2 K had the TR component smaller than the AP component at any angle. There is not any dysprosium moment orientation with such a coefficient configuration in figure \ref{Fig_coeff_ang_conang}. The same occurs at RT. In this case, the AP experimental coefficient is overwhelm higher than the PP coefficient. Moreover, the PP coefficient reaches such a value only when the dysprosium magnetic moment orientation is in the plane. All these means that the dysprosium spectra of the DCC and DCM samples can not be understood using a single distribution of dysprosium magnetic moments. 

This finding supports the conclusion derived from the low magnetic moment value of dysprosium at RT in section \ref{XMCD Dy}: the observed spectra must be the result of the combination of the spectra from, at least, two different distribution of dysprosium magnetic moments. One of them must have the dysprosium poorly magnetically coupled or even uncoupled from cobalt. For this to happen, a possibility is to have this dysprosium physically separated from the DyCo alloy by segregation at the interface with the aluminum cap layer. RE segregation is a common effect in RE-TM alloys, with the RE always located at the free surface region \cite{GdCo_segre_PRB,NdCo_EXAFS, NdCo_PRB,TbFeCo_segre_PRM}. Then, the total spectra can not be built by the simple summation of two spectra. That would be the case if the two types of dysprosium atoms were uniformly distributed all through the depth of the film. When TEY detection is used, the measured spectral intensity is build from the secondary electrons current which depends on the depth from where they are emitted. Therefore, the excitations produced at the surface are more heavily weighted than the originated in the layers underneath the segregated dysprosium. For instance, a possible way to explain the larger than expected intensity of component AP found in the spectra is by the presence of dysprosium with low magnetic moment (large cone opening angle) at the topmost surface of the alloy.

The intensity of the secondary electrons collected by the detector depends on the depth from where they are generated by two factors: their scape depth in the excited layers ($\lambda_{e}$), and the X ray intensity which is absorbed as it propagates through them. This causes distortions in the measured spectra with respect to the true absorption coefficient of the analyzed sample. These saturation effects becomes more important as the angle of incidence is increased, because the X ray penetration depth becomes similar to the secondary electrons scape depth \cite{saturation_stohr}. Such effects are taken into account when the spectra are analyzed to recover the absorption coefficients that were originated using a method which is fully described in \cite{NdCo_PRB}. Note that this analysis process assumes the presence of a single layer. 

Then, two steps are required to untangle the experimental spectra in their contributions from the different regions at different depths from the films surface. The first step is to model the experimental spectra assuming the presence of, at least, two layers with different thickness, dysprosium concentrations, magnetic moment distributions and secondary electron scape lengths. The modeling of the spectra is done using the absorption cross section of the PP, AP and TR components (i.e., the related peak shapes $A^{-1}_{JJ'}$, $A^{1}_{JJ'}$ and $A^{0}_{JJ'}$), multiplied by their corresponding angle dependent coefficients shown before. The sum of the contribution of each layer is weighted depending on their specific thickness and depth. The resulting spectral intensity follows the expression:

%\begin{widetext}
%\begin{equation}
%I(E)=\frac{I_{0n}}{\cos \theta_{i}}\left( \int^{z_{1}}_{0}\mu_{1} (E) e^{-(\frac{\mu_{1}(E)}{\cos \theta_{i}} +\frac{1}{\lambda_e1})z}dz
%+ e^{-(\frac{\mu_{1} (E)}{\cos \theta_{i}} z_1 )} \int^{\infty}_{z_1}\mu_2 (E) e^{-(\frac{\mu_2 (E)}{\cos \theta_{i}} +\frac{1}{\lambda_e2} )z}dz \right)
%\end{equation}
%\end{widetext}

\begin{widetext}
\begin{equation}
I(E)\approx\frac{I_{0n}}{\cos \theta_{i}} \left(\mu_{1}(E) \frac{(1-e^{-\frac{z_1}{\lambda_{e1}}})}{ \frac{\lambda_{e1}\mu_{1}(E)}{\cos \theta_{i}} +1}+\mu_{2} (E) \frac{\lambda_{e2}}{\lambda_{e1}} \frac{e^{-\frac{z_1}{\lambda_{e2}}}}{\frac{\lambda_{e2}\mu_{2}(E)}{\cos \theta_{i}}+1}\right) \label{eq_abs_depth}
\end{equation}
\end{widetext}

This equation has been calculated assuming that the first layer is thinner than the absorption coefficient ($z<<\mu_{1,2}(E)$), which is applicable in our case.

In the second step, the modeled spectrum passes through the same process as the actual experimental spectrum, described in \cite{NdCo_PRB} to extract what it could be called the effective absorption coefficient. This is a necessary step to compare the saturation corrected experimental effective absorption coefficient, where only a single layer is assumed in the spectrum saturation correction process, with the modeled one.

The parameters to enter in the modeling of the spectra are the thickness of the first layer, $\tau_{1}$, the dysprosium concentration $\rho_{i}$ of the top layer and the underlayer, the secondary electron scape length $\lambda_{ei}$ for each of the layers, the overall escape length used to correct saturation effects $\lambda_{ei}$, and the opening and inclination angles of the cone of each layer, $\theta_{Ci}$ and $\varphi_{Ci}$. The model that significantly best fitted the spectra required an additional parameter for each layer which represented the portion of dysprosium atoms with moments in the opposite orientation. 

%It is possible that this dysprosium atoms were mixed with aluminum of the capping layer, or even oxidized, reducing its effective density. This will have an influence in the fitted thickness of this first layer, which will be undervalued. This will affect also the concentration of the under layer since it will be more reduced than the the reduction in intensity of its contribution to the final spectra due to the Several magnetic moment distribution configurations were tested which included additional variables. 

Some assumptions were done to chose the value of some of the parameters of the model. The cobalt concentration of the first layer was set to zero in the first layer in base of the expected RE segregation.  The secondary electron scape length was set to 25 \AA\ for both layers. This is the length used in the saturation correction of the experimental spectrum, which was the one that worked the best in these alloys \cite{NdCo_PRB}. The validity of this parameter to correct the saturation effects was tested by measuring the experimental ratio between the AP and PP component extracted from the XMCD spectra. This ratio has to be constant for any field orientation and probed sample. Its value was identical to the theoretical one \cite{Goedkoop_thesis} within the experimental error. The goodness of the fits were evaluated by minimizing a $\chi^{2}$ function. Different values of the above mentioned fixed parameters were tried to check the incidence in the results of the fits. Their variation over a relatively wide range of values did not change in a substantial way the conclusions of the spectral fits here presented.

The accuracy of the fitted parameters and the discrimination accuracy of the possible configurations was increased by fitting at the same time the spectra taken at opposite circular polarizations. Both spectra are symmetrical in their spectral components and, therefore, they must use the same fitting parameters: the TR component does not change with circular polarization light and the PP and AP components becomes AP and PP components in the opposite helicity. Obviously, the condition for the fits of yielding the same magnetic moment value as the experimental one was imposed. The inadequacy of the spectra to be fitted with a single component was tested. The quality of the fits was very low in this case, with $\chi^{2}$ values two orders of magnitude higher than the resulted from the proposed models, confirming the previous conclusion extracted from the deconvolution of the spectra in their PP, AP and TR components. 

The spectral fittings were done using the PP and AP components with (corrected) and without (uncorrected) the overlapping component, using the related TR component in each case. We will only show the results corresponding to the corrected components. The results were essentially the same using the uncorrected components.

The described analysis of the spectra used two layers for the spectra modeling.  Include a larger number of layers might compromise the accuracy of the analysis due to the consequent increase in the number of variables. Also, the thickness of the effective probed layer, which is estimated to be of the order of 2 to 3 nm, leaves little room for a better in depth accuracy for the method, because the intensity is exponentially reduced with depth. The use of two layers is also justified by posterior measurements on the same samples by X ray photoemission which showed a significantly larger dysprosium concentration at the surface than the nominally expected. The next section discusses the results of the fits.

\subsection{Results and discussion of the fits}
\begin{figure}
\includegraphics[width=8 cm]{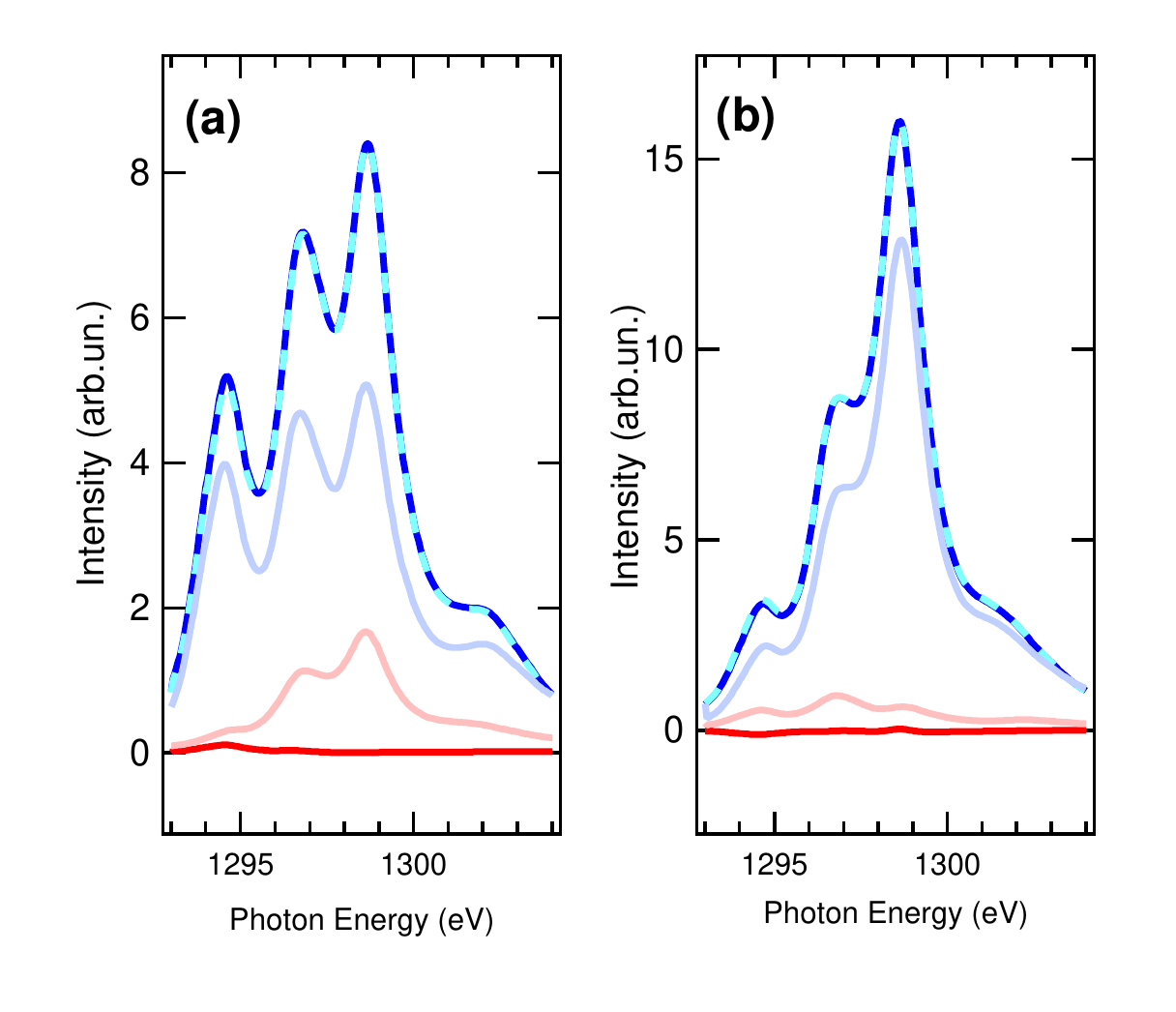}
\caption{ Fit of the circularly polarized spectra (a) C$^{+}$ and (b) C$^{-}$ of Dy $M_{5}$ of sample DCC taken at RT. The dark blue line is the experimental spectrum, the dashed light blue line is the fitted curve, and the red line is the residual. The light red line is the spectral component for the first layer and the blue line is the spectral component for the under layer. \label{fig_fit_DCC_RT_CmyCp}}
\end{figure}

\begin{figure}
\includegraphics[width=8 cm]{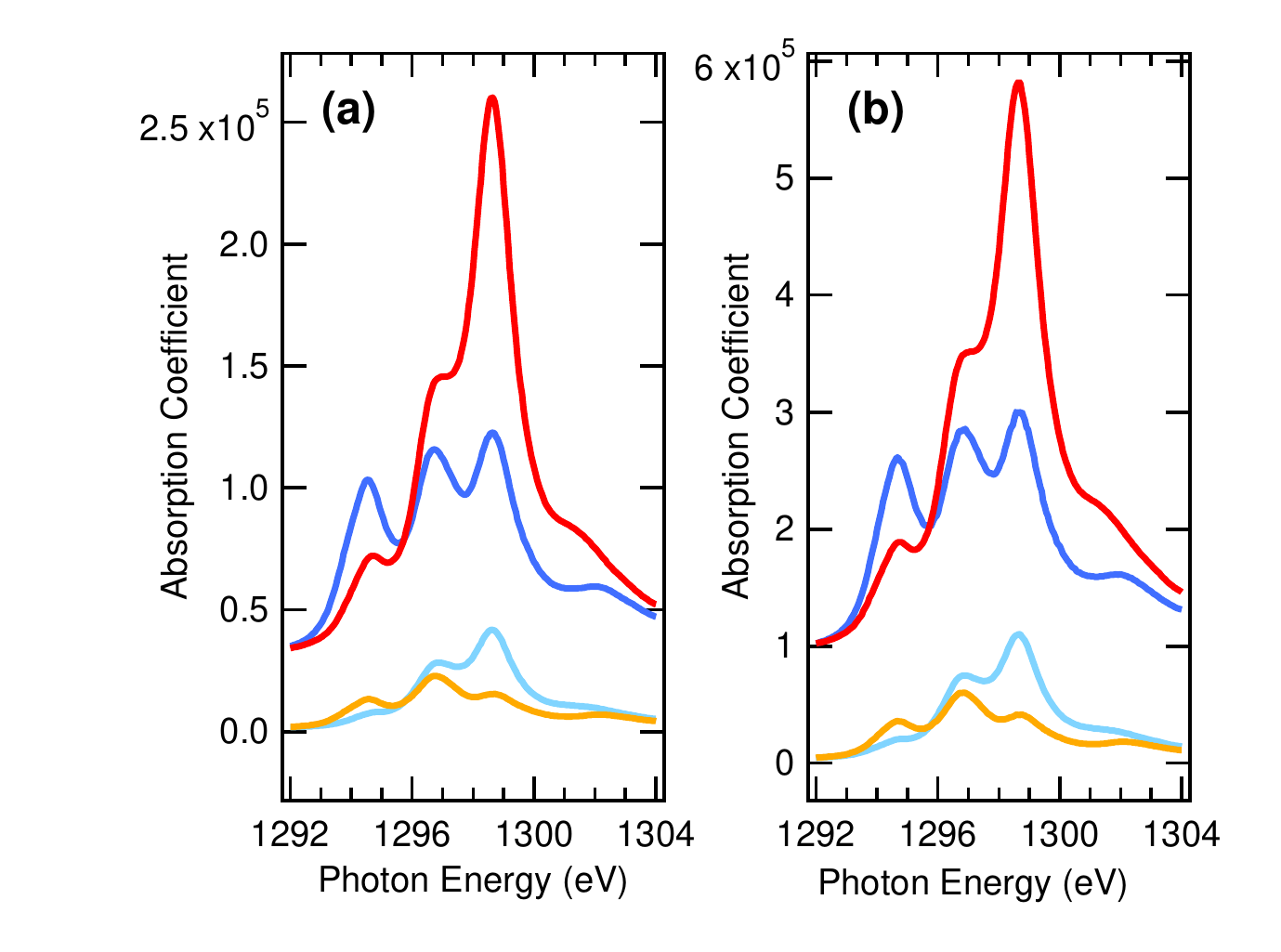}
\caption{ Espectral components of the Dy $M_{5}$ circularly polarized spectra of sample DCC taken at RT and at (a) 0$^{\circ}$ and (b) 70$^{\circ}$ orientation field angles. For C$^{+}$ polarization, blight blue line for the top layer and dark blue line for the under layer.  For C$^{-}$ polarization, blight orange line for the top layer and red line for the under layer.   \label{Fig_DyM5_L1L2_rt0y70}}
\end{figure}

Table \ref{table_fitsM5} shows the result of the fits with the lowest $\chi^{2}$ of the spectra taken at RT and 2 K. Figure \ref{fig_fit_DCC_RT_CmyCp} shows the fitting of the circularly polarized spectra of sample DCC at RT an 0$^{\circ}$ incident angle as an example of the quality of the fit.  In the model used for these fits, the top layer had its average moment direction opposite to that of the dysprosium in the under-layer. The underlayer had two dysprosium moment components that shared the same moment distribution, but having their average moment in opposite directions. Figure \ref{Fig_DyM5_L1L2_rt0y70} shows the spectra of each of the layers used to adjust to the spectra taken at RT and at 0$^{\circ}$ and 70$^{\circ}$ incident angles. The proportion of the component not AF oriented to cobalt was estimated in 25$\%$ of the total. The cone opening angle of these two components was of about 50$^{\circ}$. In the top layer, the moment orientation was distributed uniformly in a semi-sphere. These parameters did not change too much with field angle orientation or sample. The only observed changes occurred in the effective atomic cobalt concentration, which increased from 3.6 to 4.9 atoms of cobalt per 1 of dysprosium from normal to grazing incidence. This  change has the effect of increasing the spectral weight of the top layer with increasing field orientation angle in a slight higher proportion than the predicted by equation \ref{eq_abs_depth}. This effect was even stronger in the fits of the spectra taken at 2 K. The possible explanation to this will be discussed later on.

The total moment of dysprosium in the underlayer was 3.8$\mu_{B}$. It matched the expected from the VSM data (see table \ref{tab_Comp_Temp_DCC_DCM}) using the cobalt moment deduced from XMCD. It is important to notice that this value was obtained considering the presence of a significant portion of dysprosium atoms whose average magnetic moment orientation was the same as cobalt. There is no other way to obtain such a low magnetic moment in dysprosium. This is the reason why the AP component is so intense at this temperature. The presence of this phase in the underlayer indicates that it should also exists in the bulk, although in a lower proportion in base of the estimated value of its dysprosium moment in that region. The magnetic interaction of this kind of dysprosium with cobalt must be either too weak or null and it might be caused by a segregation process and/or by disorder. The relatively narrow moment distribution of this dysprosium compared to the fitted on the top layer could be explained by being the former in closer interaction with other magnetic atoms increasing the mean field felt by them, whereas the latter might be mixed with those of the capping layer at the interface. At low temperature, this type of dysprosium in the under-layer should contribute to increase the total magnetic moment of the dysprosium sublattice. The presence of these weakly coupled dysprosium atoms at the region probed by TEY would explain the apparent increase in the $T_{Comp}$ at the surface with respect to the bulk measured by XMCD using TEY detection, since its average magnetic moment is oriented in the same direction as the applied field. It should be also the cause,of the observed small decoupling between the cobalt and dysprosium hysteresis loops (see section \ref{magnetometry}).

\begin{figure}
\includegraphics[width=8 cm]{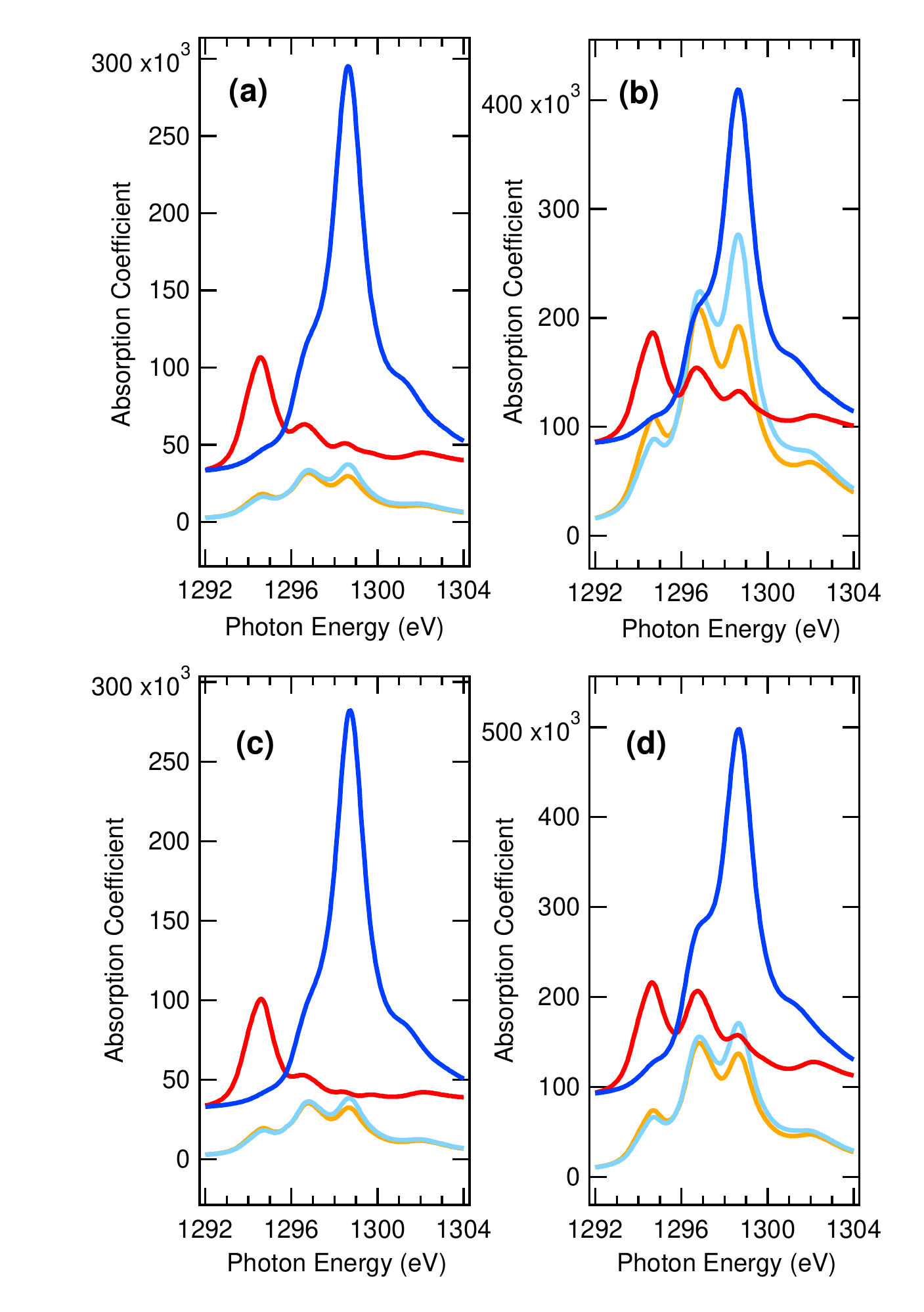}
\caption{  Espectral components of the circularly polarized spectra Dy $M_{5}$  taken at 2 K and at (a) 0$^{\circ}$ and (b) 70$^{\circ}$ orientation field angles for sample DCC, and at (c) 0$^{\circ}$ and (d) 70$^{\circ}$ for sample DCM. For C$^{+}$ polarization, blight blue line for the top layer and dark blue line for the under layer.  For C$^{-}$ polarization, blight orange line for the top layer and red line for the under layer.\label{Fig_DyM5_L1L2_0y70}}
\end{figure}

At 2 K, the underlayer has a single component, but the top layer required two types of dysprosium with opposite averaged moment directions. The moment distribution for this layer caused a reduction of its total moment to almost zero, becoming close to a magnetically dead layer. This change in the magnetic behavior of the top layer could be attributed to the presence of some oxidized dysprosium which is AF below 4 K \cite{AF_DyO}. The oxidation state does not cause any change in the Dy $M_{4,5}$ spectra \cite{DyO_XAS} but it does in the Co $L_{2,3}$ spectra. Cobalt spectra did not show any feature related to oxidized cobalt, what it could be explained by the expected location of the possibly oxidized segregated dysprosium at the interface with the aluminum cap layer. Figure \ref{Fig_DyM5_L1L2_0y70} shows the fitted components for both layers at the two circular polarizations for 0$^{\circ}$ and 70$^{\circ}$ incidence angles at 2 K, and for both samples.

\begin{table}[h]
\caption{Fitting parameters of Dy $M_{5}$ spectra of samples DCC and DCM taken at RT and 2K, and at different sample orientations with respect to the applied field: $\tau_{1}$: thickness of the top layer; $\theta_{C1}$ and $\theta_{C2}$: half opening cone angle for the fanning distribution of the magnetic moment of dysprosium in the top and under the top probed layers, respectively; $\varphi_{C2}$: cone inclination angle with respect to the beam (field) orientation in the underlayer; $m_{C2}$ estimated magnetic moment of the underlayer; $\rho_{2}$: effective atomic concentration of dysprosium at the under layer; AP$_{1}$: proportion of antiparallel dysprosium in the first layer (only at 2 K); $\chi^{2}$: statistical goodness of the fit.\label{table_fitsM5}}
\begin{ruledtabular}
\begin{tabular}{lllllll}
Sample & T & & \multicolumn{4}{l}{Field Orientation Angle } \\
\colrule\hline
&&&0$^{\circ}$&20$^{\circ}$&45$^{\circ}$&70$^{\circ}$ \\ \hline
DCC & RT &$\tau_{1}$& 1.6(1) \AA\ & - & - & 1.5(1) \AA\ \\
& &$\theta_{C1}$& 90$^{\circ}$ & -& - & 90$^{\circ}$\\
& &$\theta_{C2}$& 53.5(1)$^{\circ}$ &  -& -& 54.1(1)$^{\circ}$\\	
& &$AF_{C2}$& 0.26(1) & - &-&0.26(1) \\
& &$m_{C2}$& -3.8(1) & - &-&-3.8(1) \\
& &$\rho_{2}$& DyCo$_{3.6(2)}$  & - &-&DyCo$_{4.9(3)}$\\
& &$\chi^{2}$& 0.6 & - &-&0.8 \\
& 2 K &$\tau_{1}$&2.1(1) \AA\ & 2.0(1) \AA\ &2.8(1) \AA\ &5.6(1) \AA\ \\
& &$\theta_{C1}$& 90$^{\circ}$&90$^{\circ}$&90$^{\circ}$ &90$^{\circ}$ \\
& &$\theta_{C2}$& 35.9(1)$^{\circ}$ & 45.7(2)$^{\circ}$&47.5(2)$^{\circ}$& 40.2(1)$^{\circ}$\\	
& &$\varphi_{C2}$& 0 & 10$^{\circ}$ & 9$^{\circ}$ &47$^{\circ}$ \\
& &$m_{C2}$& 8.9 & 8.2 & 8.1 & 5.8 \\
& &$\rho_{2}$& DyCo$_{5.1}$ & DyCo$_{5.0}$ &DyCo$_{5.2}$&DyCo$_{5.8}$ \\
& &AP$_{1}$& 56$\%$ & 64$\%$ &56$\%$ &39$\%$ \\
& &$\chi^{2}$& 1.74 & 3.23 &2.0&0.9 \\ \hline
DCM & RT &$\tau_{1}$&- & - & - &1.5 (1) \AA\ \\
& &$\theta_{C1}$& - & -& - & 90$^{\circ}$ \\
& &$\theta_{C2}$& - &  -& -& -56.7(1)$^{\circ}$\\	
& &$AF_{C2}$& - & - &-&0.26(1) \\
& &$m_{C2}$& - & - &-&-3.7(1) \\
& &$\rho_{2}$& - & - &-& DyCo$_{5.3}$ \\
& &$\chi^{2}$& - & - &-&0.4 \\
& 2 K &$\tau_{1}$& 2.6(2) \AA\ & 2.4(2) \AA\ & 3.3(2) \AA\ &  3.7(2) \AA\\\
& &$\theta_{C1}$& 90$^{\circ}$& 90$^{\circ}$& 90$^{\circ}$ & 90$^{\circ}$\\
& &$\theta_{C2}$& 27.0(1)$^{\circ}$ &  32(1)$^{\circ}$& 44(1)$^{\circ}$& 48(1)$^{\circ}$\\	
& &$\varphi_{C2}$& 0(1)$^{\circ}$ & 6(1)$^{\circ}$&  25(1)$^{\circ}$&   46(1)$^{\circ}$ \\
& &$m_{C2}$& 9.3(2) & 9.0(2) &7.8(2)& 6.1(2) \\
& &$\rho_{2}$& DyCo$_{4.9(1)}$ & DyCo$_{5.3(1)}$ &DyCo$_{5.4(1)}$&DyCo$_{5.4(1)}$ \\ 
& &AP$_{1}$& 61$\%$ & 61$\%$ & 58$\%$ & 52$\%$ \\
& &$\chi^{2}$& 3.3 & 3.3 & 3 & 2.8 \\
\end{tabular}
\end{ruledtabular}
\end{table}

\begin{figure}
\includegraphics[width=8 cm]{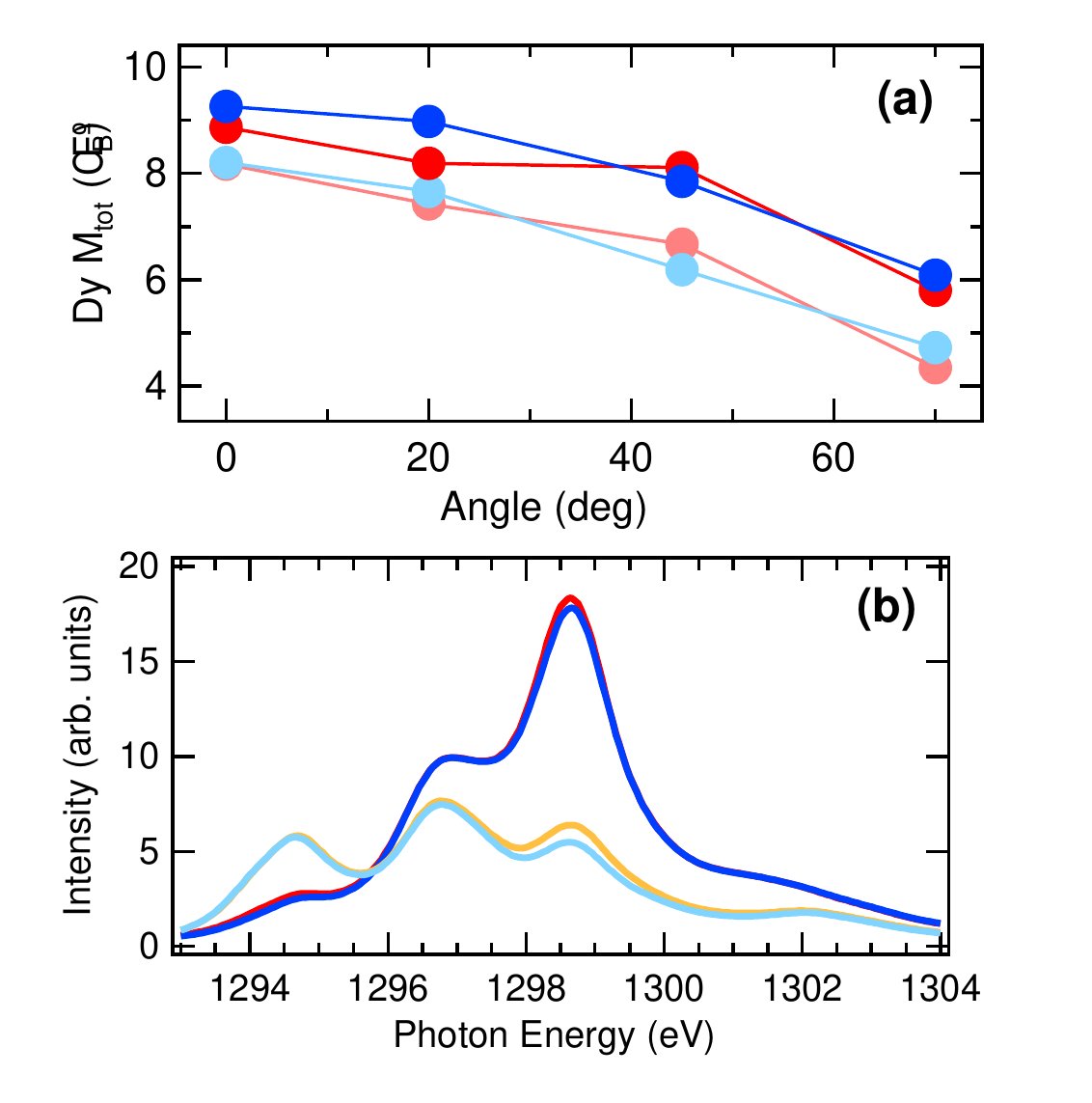}
\caption{(a) Comparison between the magnetic moment of dysprosium in the underlayer resulted from the "`two layers model" fits (red dots, sample DCC; blue dots, sample DCM), and the magnetic moments measured considering a single dysprosium magnetic state, as a function of the field orientation angle. (b) Comparison between the spectra of the Dy $M_{5}$ circularly polarized spectra of samples DCC and DCM taken at 2 K and at 70$^{\circ}$. Dark (C$^{+}$) and light (C$^{-}$) blue lines for sample DCM and red (C$^{+}$) and orange (C$^{-}$) lines for sample DCC.  \label{Fig_model}}
\end{figure}

%The thickness of the top layer was increased with respect to RT about 0.5 \AA, consequently increasing the effective concentration of the underlayer. As the angle of incidence increased, the effective thickness of the top layer increased up to about 3.8 \AA, as well as the dysprosium concentration decreased, indicating that the extension in thickness of this layer is likely larger than the estimated from the fits. These differences with respect to the parameters observed at RT could be understood considering that what it is adjust in the fits is their magnetic moment distribution not its chemical composition. From here it could be induce the presence of a gradation layer which should has a larger thickness than the deduced at RT.

Due to the low magnetization of the top layer, the values of the dysprosium moments in the under-layer deduced from the fits were larger than the extracted from the XMCD measurements (see table \ref{table_XMCD_DyCo}). They are compared in figure \ref{Fig_model} (a). The values are similar in both samples. An important consequence of the values of the cobalt and dysprosium moments found by the fits at 2 K is the expected enhancement of the magnetization at the region probed by TEY. At normal incidence, and keeping the nominal dysprosium to cobalt concentration, both samples yielded the same magnetization, 345 $emu/cm^{3}$. This is almost 5 times for sample DCM, and 3 times for DCC, their magnetization in the bulk measured by VSM at 10 K. The magnetization enhancement would be even higher if the dysprosium concentration was higher. This high magnetization of the surface region has to do mainly with the increased momentum of dysprosium with respect to the expected in the bulk. Attending to the results of the fits obtained at RT, the increment in the dysprosium total magnetic moment can be explained by the portion of the weakly magnetically exchange-coupled-to-cobalt dysprosium atoms which, at this temperature, are ferromagnetic. The magnetic anisotropy of these dysprosium atoms is expected to be lower than those exchange coupled to cobalt,  what would explain the apparent lower anisotropy of the dysprosium sublattice. Also, the anisotropy field $H_{k}$ is expected to decrease due to the increase in the magnetization.  In reference to this, the deviations of the dysprosium cone distribution with respect to the beam (angle $\varphi_{C2}$) and the increasing opening of the cone angle (angle $\theta_{C2}$) with the increassing field orientation angle $\theta_{i}$ with respect to the easy axis is the qualitatively expected behavior in the presence of an uniaxial PMA where dysprosium atoms are subject to different local anisotropy energies. The variation in the dysprosium magnetic distribution were similar in both samples (see table \ref{table_fitsM5}). A higher insight of the magnetic interactions and anisotropy of the dysprosium and cobalt atoms could be gain by calculating them in base to the observed magnetic moment distributions resulted from the fits \cite{Honorino_APL}.

The required use of dysprosium with opposite average moment directions in the fits of the spectra at RT and at 2 K is consequent with the observations done in the previous section, where it was shown that the intensity of the deconvolved AP component was too high compared to the TR component. The fits showed, however, some apparent inconsistencies in the values of the thickness of the first layer and the cobalt concentration of the underlayer that were pointed out at the beginning of this section. These consisted in an increased value of both parameters with the incident angle, indicating that the spectral weight given to the top layer in equation \ref{eq_abs_depth} was below the required in the fits. The second inconsistency was an increase in the values of these parameters when the temperature decreased to 2 K. These two effects could be attributed, at least partially, to the simplicity of the model, which does not consider any gradation in the moment distribution from one layer to the other, and to the fact that the spectra represents moment distributions, which might change with the temperature, and not chemical changes.

%The observed changes in the top layer thickness and the under-layer cobalt concentration have a regular variation with $\theta_{i}$: as this angle increases, the contribution of the top layer to the spectra increases (its thickness increases) and that of the under layer decreases (its cobalt concentration increases). Apparently, the parameters introduced in equation \ref{eq_abs_depth} underestimate the effects of the top layer thickness and concentration. 

The thickness values for the top layer used in the fits are small, of the order of one dysprosium monolayer, and its highest value is not higher than two monolayers. These values might not be the actual ones. For instance, they are below the estimated roughness of the film surfaces. Moreover, it is expected some diffusion of dysprosium within the aluminum capping layer that should extend the thickness of the top layer dysprosium, decreasing its density and concentration. These details are not contemplated in the fit, where the concentration of the top layer was fixed to pure dysprosium. Therefore, thickness and concentrations would be acting as effective parameters. This would explain the apparent increase in the thickness of the top layer and/or in cobalt concentration in the underlayer at higher incident angles. In this situation, the signal from the underlayer will be taken from deeper regions, which have lower secondary electrons emitted and a larger path for them to scape up to the surface of the sample. The final effect is an increase in the spectral weight of the top layer in the spectra. 

However, this can not fully explain the larger spectral weight of the top layer in the spectra taken at 2 K and 70$^{\circ}$ angle incidence with respect to the spectra taken at RT. This can be checked by comparing the spectral components used to fit the spectra taken at RT in figure \ref{Fig_DyM5_L1L2_rt0y70} with those taken at 2 k shown in figure \ref{Fig_DyM5_L1L2_0y70}. Moreover, the spectral weight of the top layer is different depending on the sample, being specially large in sample DCC, as it can be observed in figure \ref{Fig_DyM5_L1L2_0y70}. The difference of the fitted parameters in these two samples at 70$^{\circ}$ incidence angle is related to their different intensity in their AP component, as it is shown in figure \ref{Fig_model} (b). 

At 2 K, a forced reduction in the thickness of the first layer in the fits of the spectra taken at high angle of incidence to match those at low incidence angles was possible but the fits yielded a reduction in the magnetization of the under layer. The related magnetic configurations had to be discard because the yielded dysprosium moments for the underlayer were below the deduced from XMCD (table \ref{table_XMCD_DyCo}). Moreover, such a low magnetic moment would be inconsistent with the values obtained at lower angles. If these changes were related to an irregular distribution of segregated dysprosium across the film, their lateral size should be in the range of the size of the beam which could be in this case of the order of 300 $\mu$m. 

A plausible explanation for this effect might have to do with the sensitivity of the dysprosium spectrum to the moment distribution across the thickness of the film and not to its chemistry. The magnetization of the probed dysprosium increases from RT to 2 K. Pure dysprosium becomes ferromagnetic at temperatures below 80 K and oxidized dysprosium is AF at temperatures below 4 K. This means that the changes in the orientation of the dysprosium moment should be more gradual from the top layer to the under-layer at 2 K than at RT, as the fits suggests. As the orientation of the applied magnetic field deviates from the easy axis, the top layer with its nearly zero average magnetic momentum would penetrate into the film driven by the increasing energy of the anisotropy field that reduced the magnetization in the direction of the applied field \cite{reflectivity_AF_F}. More measurements done at different temperatures are needed to fully test this explanation, what will definitively prove the high sensitivity of the proposed analysis method to the variations in the magnetic moment of the REs at the region probed by TEY.

\section{Conclusions}

To summarize, the magnetism of the dysprosium and the cobalt sublattices in ferromagnetic DyCo thin films with PMA has been investigated by XMCD spectroscopy using TEY detection. Some unexpected results were observed in the magnetic behavior of the dysprosium sublattice. The measurements done at the Co $L_{2,3}$ showed that the PMA of the films must stem from the RE sublattice. However, at 2 K, when the PMA energy of the films was the highest, the probed cobalt sublattice resulted to be magnetically more anisotropic than the dysprosium sublattice. Additionally, the magnetic moment of the dysprosium sublattice measured at RT was lower than expected if all moments were AF coupled to cobalt. To understand the cause of these apparent anomalies, a method to analyze the circularly polarized Dy M$_{5}$ spectra obtained by TEY has been presented, which is based in the deconvolution of their parallel, antiparallel and transverse to $J_{z}$ spectral components. This is the first time that this kind of analysis is applied in this kind of systems. This spectral analysis reveals the presence of a relatively large proportion of dysprosium with its average moment oriented in the same direction than cobalt, and a thin layer of segregated dysprosium at the top of the alloy which should be not exchanged coupled to cobalt. Some of the dysprosium in this layer is AF coupled at 2 K, suggesting that either it is partially oxidized or there exist a proportion of dysprosium which is magnetically dead at this temperature. The apparently lower magnetic anisotropy of dysprosium with the field orientation angle is explained by the thickness of the segregated dysprosium layer at the region near the surface.  A similar kind of measurements should be conducted using bulk detection sensitivity to determine the precise distribution of RE sites that contribute to the PMA of the alloy. This seems mandatory to be able to predict with accuracy the possible magnetic configurations resulting from the interaction between RE and TM sublattices.

The presented results show that, if the RE is segregated at the surface, care must be taken to remove the effect that it causes in the value of the RE magnetic moment measured using TEY detection. This affects, for instance, to the $T_{Comp}$ values of RE-TM alloys measured at the surface in this way, which could be overestimated.

The proposed analysis method can serve to detect the presence of RE in different magnetic states, either in the surface or in the bulk. The method can be extended to the circularly polarized spectra of other RE at the $M_{4,5}$ edge, whenever the angular moment $J$ of its $4f$ orbital is different of zero (i.e.,it can not be applied to Gd$^{+3}$). In those other RE, the identification of the PP, AP and TR spectral components requires a different methodology than the presented here for dysprosium since their overlap is much important. This might not be a problem because these components are well approximated by theory. They can be also isolated by spectroscopic methods using circular but also linear polarized spectroscopy at the related RE $M_{4,5}$ edges. Nevertheless, more additional work is needed to improve and test the accuracy of the method by better characterizing the region probed with other surface sensitive techniques. 

The experiment confirms the presence of segregated RE reported in other RE-TM systems. The preparation method of the alloy did not produced substantial differences in the magnetism of the cobalt and dysprosium sublattices at the surface of the alloy, indicating that the origin of their different macroscopic behaviors has to be found in the bulk or they are caused by their different microstructure. The no-dependence of the RE segregation with the deposition method (co-sputtering and alternate deposition) remarks that the segregated RE is an immediate post deposition process, which can proceed during time as it has been observed in GdCo \cite{GdCo_segre_PRB}. This is also demonstrated in the DCM film whose last deposited layer, before depositing the aluminum capping layer, was cobalt. This means that RE atoms could diffuse also in the bulk at defects like voids, cracks and grain borders within the bulk. This might affect magnetic properties like, for instance, the coercive field of the thin film since it would contribute to the formation of domain wall pinning defects. 

%Method to determine the possible different dysprosium magnetic moment distributions in the alloy. 
%Dysprosium segregation. How to avoid it. Learn what drives it
%Confirmed RE is at the origin of the anisotropy. TEY is not thge best way to unserstand the precise mechanism. It requires a bulky technique like transmission. The method might help %to determine possible microscopic inhomogeneities in the bulk as well.

\appendix
\section{Orientation dependence of light absorption\label{ApA}}
Here, it is shown the details of the calculation of the intensity absorbed by circularly polarized light when the orientation of the angular moment of the $4f$ orbital forms an angle  $\theta$ with the direction of the wavevector $\vec{k}$ which is chosen parallel to the $z$ axis. The components of the electric field of the circularly polarized light in the frame of the $4f$ angular moment, with its $z_{m}$ axis along its quantization axis, are calculated by making a first rotation of an angle $\varphi$ in the $xy$ polarization  plane ($z$ axis as the rotation axis) followed by a rotation of an angle $\theta$ around the resulted $y$ axis after the first rotation.

\begin{equation}
\left(\begin{array}{c}
\epsilon_{x_{m}}\\
\epsilon_{y_{m}}\\
\epsilon_{z_{m}}\\
\end{array}\right)=\left(\begin{array}{ccc}
\cos\varphi & -\sin\varphi & 0\\
\sin\varphi & \cos\varphi & 0\\
0 & 0 & 1\\
\end{array}\right)\left(\begin{array}{ccc}
\cos\theta & 0 & -\sin\theta \\
0 & 1 & 0\\
\sin\theta & 0 & \cos\theta\\
\end{array}\right)\left(\begin{array}{c}
\epsilon_{x}\\
\epsilon_{y}\\
\epsilon_{z}\\
\end{array}\right)
\end{equation}

Therefore, for right ($-$) and left ($+$) circular polarization:
\begin{equation}
\left(\begin{array}{c}
\epsilon_{x_{m}}\\
\epsilon_{y_{m}}\\
\epsilon_{z_{m}}\\
\end{array}\right)=\frac{\epsilon e^{\pm i\varphi}}{\sqrt{2}}\left(\begin{array}{c}
\cos\theta\\
\mp i\\
\sin\theta\\
\end{array}\right)
\end{equation}

By using this result, the electric dipole operator $P$ written in the frame of the 4$f$ orbital reads:

\begin{equation}
P=\frac{e^{\pm i\varphi}}{\sqrt{2}}\left[P^{0}_{x_{m}}\cos\theta \mp iP^{0}_{y_{m}}+P^{0}_{z_{m}}\sin\theta\right]
\end{equation}

The superscript in the dipole operator indicates the polarization state of the electric field in the frame of the 4$f$ orbital: $0$ is lineal, $1$  circular left handed and $+1$  circular right handed. The electric dipole operator can be expressed in the components parallel to the quantization axis $z_{m}$ of the 4$f$ orbital using the relations:
\begin{equation}
P^{0}_{x_{m}}=\frac{1}{\sqrt{2}}\left[P^{-1}_{z_{m}}+P^{1}_{z_{m}}\right]
\end{equation}

and 
 
\begin{equation}
P^{0}_{y_{m}}=\frac{i}{\sqrt{2}}\left[P^{-1}_{z_{m}}-P^{1}_{z_{m}}\right]
\end{equation}

Then

\begin{equation}
P=\frac{e^{\pm i\varphi}}{\sqrt{2}}\left[\frac{(\cos \theta \pm 1)}{\sqrt{2}}P^{-1}_{z_{m}}+\frac{(\cos \theta \mp 1)}{\sqrt{2}}P^{1}_{z_{m}}+P^{0}_{z_{m}}\sin \theta\right]
\label{eq_adip1}
\end{equation}

The absorption cross section for the excitation from a state $\left|\alpha m J\right\rangle$ to a state $\left|\alpha' m' J'\right\rangle$ is, in the dipole approximation:
\begin{equation}
\sigma_{\alpha m J\longrightarrow \alpha' m' J'}=4\pi^{2}\alpha_{0}\hbar\omega\left|\left\langle \alpha m J\right|P\left|\alpha' m' J'\right\rangle\right|^{2}
\label{eq_adip2}
\end{equation}
The cross section for circular polarized light is obtained substituting equation \ref{eq_dip1} in equation \ref{eq_dip2}:

\begin{widetext}
\begin{equation}
\sigma_{\alpha m J\longrightarrow \alpha' m' J'}=4\pi^{2}\alpha_{0}\hbar \omega S_{\alpha J \alpha'J'}\left[\frac{(\cos \theta \pm 1)^{2}}{4}A^{-1}_{JJ'}+\frac{(\cos \theta \mp 1)^{2}}{4}A^{1}_{JJ'}+\frac{\sin \theta^{2}}{2}A^{0}_{JJ'}\right]
\label{eq_adip3}
\end{equation}
\end{widetext}
where $S_{\alpha J \alpha'J'}$ is the radial integer and $A^{q}_{JJ'}$ are the angular integers. Under the assumption of a weak crystal field, as it happens in dysprosium, the values of the radial and angular integers do not change. Only the relative orientation of the moment with respect to incident beam changes. We redefine the angular integers as the absorption cross sections for the corresponding situation in which the polarized beam and the magnetization are parallel($q=0,\pm1$). Then, the $\theta$ dependent functions that multiply to each of the related $q=0,\pm1$ cross sections are the coefficients for the TR ($q=0$), PP ($q=-1$) and AP ($q=+1$) components:
\begin{widetext}
\begin{equation}
PP+AP+TR=\left[\frac{(\cos \theta \pm 1)^{2}}{4}A^{-1}_{JJ'}+\frac{(\cos \theta \mp 1)^{2}}{4}A^{1}_{JJ'}+\frac{\sin^{2}\theta}{2}A^{0}_{JJ'}\right]
\label{eq_adip4}
\end{equation}
\end{widetext}

\begin{acknowledgments}
J.D. acknowledges M. Valvidares for BOREAS beam line set up and his help in the beamline during data acquisition, and C. Quir\'os, and J. M. Alameda for providing the samples and the VSM measurements.This project has been supported by Spanish MINECO under grant FIS2016-76058 (AEI/FEDER, EU) and grant PID2019-104604RB/AEI/10.13039/501100011033.  
\end{acknowledgments}

\bibliography{DyCo_M5Analysis_v4.bib}

\begin{thebibliography}{47}%
\makeatletter
\providecommand \@ifxundefined [1]{%
 \@ifx{#1\undefined}
}%
\providecommand \@ifnum [1]{%
 \ifnum #1\expandafter \@firstoftwo
 \else \expandafter \@secondoftwo
 \fi
}%
\providecommand \@ifx [1]{%
 \ifx #1\expandafter \@firstoftwo
 \else \expandafter \@secondoftwo
 \fi
}%
\providecommand \natexlab [1]{#1}%
\providecommand \enquote  [1]{``#1''}%
\providecommand \bibnamefont  [1]{#1}%
\providecommand \bibfnamefont [1]{#1}%
\providecommand \citenamefont [1]{#1}%
\providecommand \href@noop [0]{\@secondoftwo}%
\providecommand \href [0]{\begingroup \@sanitize@url \@href}%
\providecommand \@href[1]{\@@startlink{#1}\@@href}%
\providecommand \@@href[1]{\endgroup#1\@@endlink}%
\providecommand \@sanitize@url [0]{\catcode `\\12\catcode `\$12\catcode
  `\&12\catcode `\#12\catcode `\^12\catcode `\_12\catcode `\%12\relax}%
\providecommand \@@startlink[1]{}%
\providecommand \@@endlink[0]{}%
\providecommand \url  [0]{\begingroup\@sanitize@url \@url }%
\providecommand \@url [1]{\endgroup\@href {#1}{\urlprefix }}%
\providecommand \urlprefix  [0]{URL }%
\providecommand \Eprint [0]{\href }%
\providecommand \doibase [0]{https://doi.org/}%
\providecommand \selectlanguage [0]{\@gobble}%
\providecommand \bibinfo  [0]{\@secondoftwo}%
\providecommand \bibfield  [0]{\@secondoftwo}%
\providecommand \translation [1]{[#1]}%
\providecommand \BibitemOpen [0]{}%
\providecommand \bibitemStop [0]{}%
\providecommand \bibitemNoStop [0]{.\EOS\space}%
\providecommand \EOS [0]{\spacefactor3000\relax}%
\providecommand \BibitemShut  [1]{\csname bibitem#1\endcsname}%
\let\auto@bib@innerbib\@empty
%</preamble>
\bibitem [{\citenamefont {Buschow}(1977)}]{buschow_RMreview77}%
  \BibitemOpen
  \bibfield  {author} {\bibinfo {author} {\bibfnamefont {K.~H.~J.}\
  \bibnamefont {Buschow}},\ }\bibfield  {title} {\bibinfo {title}
  {Intermetallic compounds of rare-earth and 3d transition metals},\
  }\href@noop {} {\bibfield  {journal} {\bibinfo  {journal} {Reports on
  Progress in Physics}\ }\textbf {\bibinfo {volume} {40}},\ \bibinfo {pages}
  {1179} (\bibinfo {year} {1977})}\BibitemShut {NoStop}%
\bibitem [{\citenamefont {Skomski}\ and\ \citenamefont
  {Sellmyer}(2009)}]{SKOMSKI_AniRE}%
  \BibitemOpen
  \bibfield  {author} {\bibinfo {author} {\bibfnamefont {R.}~\bibnamefont
  {Skomski}}\ and\ \bibinfo {author} {\bibfnamefont {D.}~\bibnamefont
  {Sellmyer}},\ }\bibfield  {title} {\bibinfo {title} {Anisotropy of rare-earth
  magnets},\ }\href
  {https://doi.org/https://doi.org/10.1016/S1002-0721(08)60314-2} {\bibfield
  {journal} {\bibinfo  {journal} {Journal of Rare Earths}\ }\textbf {\bibinfo
  {volume} {27}},\ \bibinfo {pages} {675 } (\bibinfo {year}
  {2009})}\BibitemShut {NoStop}%
\bibitem [{\citenamefont {Blanco-Rold\'an}\ \emph
  {et~al.}(2015{\natexlab{a}})\citenamefont {Blanco-Rold\'an}, \citenamefont
  {Choi}, \citenamefont {Quir\'os}, \citenamefont {Valvidares}, \citenamefont
  {Zarate}, \citenamefont {V\'elez}, \citenamefont {Alameda}, \citenamefont
  {Haskel},\ and\ \citenamefont {Mart\'{\i}n}}]{GdCo_Cristina}%
  \BibitemOpen
  \bibfield  {author} {\bibinfo {author} {\bibfnamefont {C.}~\bibnamefont
  {Blanco-Rold\'an}}, \bibinfo {author} {\bibfnamefont {Y.}~\bibnamefont
  {Choi}}, \bibinfo {author} {\bibfnamefont {C.}~\bibnamefont {Quir\'os}},
  \bibinfo {author} {\bibfnamefont {S.~M.}\ \bibnamefont {Valvidares}},
  \bibinfo {author} {\bibfnamefont {R.}~\bibnamefont {Zarate}}, \bibinfo
  {author} {\bibfnamefont {M.}~\bibnamefont {V\'elez}}, \bibinfo {author}
  {\bibfnamefont {J.~M.}\ \bibnamefont {Alameda}}, \bibinfo {author}
  {\bibfnamefont {D.}~\bibnamefont {Haskel}},\ and\ \bibinfo {author}
  {\bibfnamefont {J.~I.}\ \bibnamefont {Mart\'{\i}n}},\ }\bibfield  {title}
  {\bibinfo {title} {Tuning interfacial domain walls in gdco/gd/gdco spring
  magnets},\ }\href {https://doi.org/10.1103/PhysRevB.92.224433} {\bibfield
  {journal} {\bibinfo  {journal} {Phys. Rev. B}\ }\textbf {\bibinfo {volume}
  {92}},\ \bibinfo {pages} {224433} (\bibinfo {year}
  {2015}{\natexlab{a}})}\BibitemShut {NoStop}%
\bibitem [{\citenamefont {Morales}\ \emph {et~al.}(2004)\citenamefont
  {Morales}, \citenamefont {Mart\'{\i}n},\ and\ \citenamefont
  {Alameda}}]{GdCo_Rafa}%
  \BibitemOpen
  \bibfield  {author} {\bibinfo {author} {\bibfnamefont {R.}~\bibnamefont
  {Morales}}, \bibinfo {author} {\bibfnamefont {J.~I.}\ \bibnamefont
  {Mart\'{\i}n}},\ and\ \bibinfo {author} {\bibfnamefont {J.~M.}\ \bibnamefont
  {Alameda}},\ }\bibfield  {title} {\bibinfo {title} {Domain walls and
  macroscopic spin-flip-like metamagnetism in
  ${\mathrm{gd}}_{x}{\mathrm{co}}_{1\ensuremath{-}x}/{\mathrm{gd}}_{y}{\mathrm{co}}_{1\ensuremath{-}y}$
  exchange-coupled double layers},\ }\href
  {https://doi.org/10.1103/PhysRevB.70.174440} {\bibfield  {journal} {\bibinfo
  {journal} {Phys. Rev. B}\ }\textbf {\bibinfo {volume} {70}},\ \bibinfo
  {pages} {174440} (\bibinfo {year} {2004})}\BibitemShut {NoStop}%
\bibitem [{\citenamefont {Chen}\ \emph
  {et~al.}(2015{\natexlab{a}})\citenamefont {Chen}, \citenamefont {Lott},
  \citenamefont {Radu}, \citenamefont {Choueikani}, \citenamefont {Otero},\
  and\ \citenamefont {O}}]{SciRep_2015}%
  \BibitemOpen
  \bibfield  {author} {\bibinfo {author} {\bibfnamefont {K.}~\bibnamefont
  {Chen}}, \bibinfo {author} {\bibfnamefont {D.}~\bibnamefont {Lott}}, \bibinfo
  {author} {\bibfnamefont {F.}~\bibnamefont {Radu}}, \bibinfo {author}
  {\bibfnamefont {F.}~\bibnamefont {Choueikani}}, \bibinfo {author}
  {\bibfnamefont {E.}~\bibnamefont {Otero}},\ and\ \bibinfo {author}
  {\bibfnamefont {P.}~\bibnamefont {O}},\ }\bibfield  {title} {\bibinfo {title}
  {Observation of an atomic exchange bias effect in dyco4 film},\ }\href
  {https://doi.org/10.1038/srep18377} {\bibfield  {journal} {\bibinfo
  {journal} {Scientific Reports}\ }\textbf {\bibinfo {volume} {5}},\ \bibinfo
  {pages} {18377} (\bibinfo {year} {2015}{\natexlab{a}})}\BibitemShut {NoStop}%
\bibitem [{\citenamefont {Blanco-Rold\'an}\ \emph
  {et~al.}(2015{\natexlab{b}})\citenamefont {Blanco-Rold\'an}, \citenamefont
  {Quir\'os}, \citenamefont {Sorrentino}, \citenamefont {Hierro-Rodr\'iguez},
  \citenamefont {\'Alvarez-Prado}, \citenamefont {Valc\'arcel}, \citenamefont
  {Duch}, \citenamefont {Torras}, \citenamefont {Esteve}, \citenamefont
  {Mart\'{\i}n}, \citenamefont {V\'elez}, \citenamefont {Alameda},
  \citenamefont {E.},\ and\ \citenamefont {Ferrer}}]{CrisNat_2015}%
  \BibitemOpen
  \bibfield  {author} {\bibinfo {author} {\bibfnamefont {C.}~\bibnamefont
  {Blanco-Rold\'an}}, \bibinfo {author} {\bibfnamefont {C.}~\bibnamefont
  {Quir\'os}}, \bibinfo {author} {\bibfnamefont {A.}~\bibnamefont
  {Sorrentino}}, \bibinfo {author} {\bibfnamefont {A.}~\bibnamefont
  {Hierro-Rodr\'iguez}}, \bibinfo {author} {\bibfnamefont {L.~M.}\ \bibnamefont
  {\'Alvarez-Prado}}, \bibinfo {author} {\bibfnamefont {R.}~\bibnamefont
  {Valc\'arcel}}, \bibinfo {author} {\bibfnamefont {M.}~\bibnamefont {Duch}},
  \bibinfo {author} {\bibfnamefont {N.}~\bibnamefont {Torras}}, \bibinfo
  {author} {\bibfnamefont {J.}~\bibnamefont {Esteve}}, \bibinfo {author}
  {\bibfnamefont {J.~I.}\ \bibnamefont {Mart\'{\i}n}}, \bibinfo {author}
  {\bibfnamefont {M.}~\bibnamefont {V\'elez}}, \bibinfo {author} {\bibfnamefont
  {J.~M.}\ \bibnamefont {Alameda}}, \bibinfo {author} {\bibfnamefont
  {P.}~\bibnamefont {E.}},\ and\ \bibinfo {author} {\bibfnamefont
  {S.}~\bibnamefont {Ferrer}},\ }\bibfield  {title} {\bibinfo {title}
  {Nanoscale imaging of buried topological defects with quantitative x-ray
  magnetic microscopy},\ }\href {https://doi.org/10.1038/ncomms9196} {\bibfield
   {journal} {\bibinfo  {journal} {Nat. Commun.}\ }\textbf {\bibinfo {volume}
  {6}},\ \bibinfo {pages} {8196} (\bibinfo {year}
  {2015}{\natexlab{b}})}\BibitemShut {NoStop}%
\bibitem [{\citenamefont {Caretta}\ \emph {et~al.}(2018)\citenamefont
  {Caretta}, \citenamefont {Mann}, \citenamefont {B\"uttner}, \citenamefont
  {Ueda}, \citenamefont {Pfau}, \citenamefont {G\"unther}, \citenamefont
  {Hessing}, \citenamefont {Churikova}, \citenamefont {Klose}, \citenamefont
  {Schneider}, \citenamefont {Engel}, \citenamefont {Marcus}, \citenamefont
  {Bono}, \citenamefont {Bagschik}, \citenamefont {Eisebitt},\ and\
  \citenamefont {Beach}}]{GdCo_Skyrmions}%
  \BibitemOpen
  \bibfield  {author} {\bibinfo {author} {\bibfnamefont {L.}~\bibnamefont
  {Caretta}}, \bibinfo {author} {\bibfnamefont {M.}~\bibnamefont {Mann}},
  \bibinfo {author} {\bibfnamefont {F.}~\bibnamefont {B\"uttner}}, \bibinfo
  {author} {\bibfnamefont {K.}~\bibnamefont {Ueda}}, \bibinfo {author}
  {\bibfnamefont {B.}~\bibnamefont {Pfau}}, \bibinfo {author} {\bibfnamefont
  {C.~M.}\ \bibnamefont {G\"unther}}, \bibinfo {author} {\bibfnamefont
  {P.}~\bibnamefont {Hessing}}, \bibinfo {author} {\bibfnamefont
  {A.}~\bibnamefont {Churikova}}, \bibinfo {author} {\bibfnamefont
  {C.}~\bibnamefont {Klose}}, \bibinfo {author} {\bibfnamefont
  {M.}~\bibnamefont {Schneider}}, \bibinfo {author} {\bibfnamefont
  {D.}~\bibnamefont {Engel}}, \bibinfo {author} {\bibfnamefont
  {C.}~\bibnamefont {Marcus}}, \bibinfo {author} {\bibfnamefont
  {D.}~\bibnamefont {Bono}}, \bibinfo {author} {\bibfnamefont {K.}~\bibnamefont
  {Bagschik}}, \bibinfo {author} {\bibfnamefont {S.}~\bibnamefont {Eisebitt}},\
  and\ \bibinfo {author} {\bibfnamefont {G.~S.~D.}\ \bibnamefont {Beach}},\
  }\bibfield  {title} {\bibinfo {title} {Fast current-driven domain walls and
  small skyrmions in a compensated ferrimagnet},\ }\href
  {https://doi.org/10.1038/s41565-018-0255-3} {\bibfield  {journal} {\bibinfo
  {journal} {Nature Nanotech.}\ }\textbf {\bibinfo {volume} {13}},\ \bibinfo
  {pages} {1154} (\bibinfo {year} {2018})}\BibitemShut {NoStop}%
\bibitem [{\citenamefont {Hierro-Rodr\'{\i}guez}\ \emph
  {et~al.}(2020)\citenamefont {Hierro-Rodr\'{\i}guez}, \citenamefont
  {Quir\'os}, \citenamefont {Sorrentino}, \citenamefont {Alvarez-Prado},
  \citenamefont {Mart\'{\i}n}, \citenamefont {Alameda}, \citenamefont
  {McVitie}, \citenamefont {Pereiro}, \citenamefont {V\'elez},\ and\
  \citenamefont {Ferrer}}]{AureNat_2020}%
  \BibitemOpen
  \bibfield  {author} {\bibinfo {author} {\bibfnamefont {A.}~\bibnamefont
  {Hierro-Rodr\'{\i}guez}}, \bibinfo {author} {\bibfnamefont {C.}~\bibnamefont
  {Quir\'os}}, \bibinfo {author} {\bibfnamefont {A.}~\bibnamefont
  {Sorrentino}}, \bibinfo {author} {\bibfnamefont {L.~M.}\ \bibnamefont
  {Alvarez-Prado}}, \bibinfo {author} {\bibfnamefont {J.~I.}\ \bibnamefont
  {Mart\'{\i}n}}, \bibinfo {author} {\bibfnamefont {J.~M.}\ \bibnamefont
  {Alameda}}, \bibinfo {author} {\bibfnamefont {S.}~\bibnamefont {McVitie}},
  \bibinfo {author} {\bibfnamefont {E.}~\bibnamefont {Pereiro}}, \bibinfo
  {author} {\bibfnamefont {M.}~\bibnamefont {V\'elez}},\ and\ \bibinfo {author}
  {\bibfnamefont {S.}~\bibnamefont {Ferrer}},\ }\bibfield  {title} {\bibinfo
  {title} {Revealing 3d magnetization of thin films with soft x-ray tomography:
  magnetic singularities and topological charges},\ }\href
  {https://doi.org/10.1038/s41467-020-20119-x} {\bibfield  {journal} {\bibinfo
  {journal} {Nat. Commun.}\ }\textbf {\bibinfo {volume} {11}},\ \bibinfo
  {pages} {6382} (\bibinfo {year} {2020})}\BibitemShut {NoStop}%
\bibitem [{\citenamefont {Mark\'o}\ \emph {et~al.}(2019)\citenamefont
  {Mark\'o}, \citenamefont {Vald\'es-Bango}, \citenamefont {Quir\'os},
  \citenamefont {Hierro-Rodr\'{\i}guez}, \citenamefont {Vélez}, \citenamefont
  {Mart\'{\i}n}, \citenamefont {Alameda}, \citenamefont {Schmool},\ and\
  \citenamefont {Alvarez-Prado}}]{Luis_APL}%
  \BibitemOpen
  \bibfield  {author} {\bibinfo {author} {\bibfnamefont {D.}~\bibnamefont
  {Mark\'o}}, \bibinfo {author} {\bibfnamefont {F.}~\bibnamefont
  {Vald\'es-Bango}}, \bibinfo {author} {\bibfnamefont {C.}~\bibnamefont
  {Quir\'os}}, \bibinfo {author} {\bibfnamefont {A.}~\bibnamefont
  {Hierro-Rodr\'{\i}guez}}, \bibinfo {author} {\bibfnamefont {M.}~\bibnamefont
  {Vélez}}, \bibinfo {author} {\bibfnamefont {J.~I.}\ \bibnamefont
  {Mart\'{\i}n}}, \bibinfo {author} {\bibfnamefont {J.~M.}\ \bibnamefont
  {Alameda}}, \bibinfo {author} {\bibfnamefont {D.~S.}\ \bibnamefont
  {Schmool}},\ and\ \bibinfo {author} {\bibfnamefont {L.~M.}\ \bibnamefont
  {Alvarez-Prado}},\ }\bibfield  {title} {\bibinfo {title} {Tunable
  ferromagnetic resonance in coupled trilayers with crossed in-plane and
  perpendicular magnetic anisotropies},\ }\href
  {https://doi.org/10.1063/1.5104341} {\bibfield  {journal} {\bibinfo
  {journal} {Applied Physics Letters}\ }\textbf {\bibinfo {volume} {115}},\
  \bibinfo {pages} {082401} (\bibinfo {year} {2019})}\BibitemShut {NoStop}%
\bibitem [{\citenamefont {Mangin}\ \emph {et~al.}(2014)\citenamefont {Mangin},
  \citenamefont {Gottwald},\ and\ \citenamefont
  {Lambert}}]{2014_enginering_nature}%
  \BibitemOpen
  \bibfield  {author} {\bibinfo {author} {\bibfnamefont {S.}~\bibnamefont
  {Mangin}}, \bibinfo {author} {\bibfnamefont {M.}~\bibnamefont {Gottwald}},\
  and\ \bibinfo {author} {\bibfnamefont {C.~e.~a.}\ \bibnamefont {Lambert}},\
  }\bibfield  {title} {\bibinfo {title} {Engineered materials for all-optical
  helicity-dependent magnetic switching},\ }\href
  {https://doi.org/10.1038/nmat3864} {\bibfield  {journal} {\bibinfo  {journal}
  {Nat. Mater.}\ }\textbf {\bibinfo {volume} {13}},\ \bibinfo {pages} {286}
  (\bibinfo {year} {2014})}\BibitemShut {NoStop}%
\bibitem [{\citenamefont {Schubert}\ \emph {et~al.}(2014)\citenamefont
  {Schubert}, \citenamefont {Hassdenteufel}, \citenamefont {Matthes},
  \citenamefont {Schmidt}, \citenamefont {Helm}, \citenamefont {Bratschitsch},\
  and\ \citenamefont {Albrecht}}]{2014_All_Optical_APL}%
  \BibitemOpen
  \bibfield  {author} {\bibinfo {author} {\bibfnamefont {C.}~\bibnamefont
  {Schubert}}, \bibinfo {author} {\bibfnamefont {A.}~\bibnamefont
  {Hassdenteufel}}, \bibinfo {author} {\bibfnamefont {P.}~\bibnamefont
  {Matthes}}, \bibinfo {author} {\bibfnamefont {J.}~\bibnamefont {Schmidt}},
  \bibinfo {author} {\bibfnamefont {M.}~\bibnamefont {Helm}}, \bibinfo {author}
  {\bibfnamefont {R.}~\bibnamefont {Bratschitsch}},\ and\ \bibinfo {author}
  {\bibfnamefont {M.}~\bibnamefont {Albrecht}},\ }\bibfield  {title} {\bibinfo
  {title} {All-optical helicity dependent magnetic switching in an artificial
  zero moment magnet},\ }\href {https://doi.org/10.1063/1.4866803} {\bibfield
  {journal} {\bibinfo  {journal} {Applied Physics Letters}\ }\textbf {\bibinfo
  {volume} {104}},\ \bibinfo {pages} {082406} (\bibinfo {year}
  {2014})}\BibitemShut {NoStop}%
\bibitem [{\citenamefont {Becker}\ \emph {et~al.}(2017)\citenamefont {Becker},
  \citenamefont {Tsukamoto}, \citenamefont {Kirilyuk}, \citenamefont {Maan},
  \citenamefont {Rasing}, \citenamefont {Christianen},\ and\ \citenamefont
  {Kimel}}]{SpinFlop_PRL}%
  \BibitemOpen
  \bibfield  {author} {\bibinfo {author} {\bibfnamefont {J.}~\bibnamefont
  {Becker}}, \bibinfo {author} {\bibfnamefont {A.}~\bibnamefont {Tsukamoto}},
  \bibinfo {author} {\bibfnamefont {A.}~\bibnamefont {Kirilyuk}}, \bibinfo
  {author} {\bibfnamefont {J.~C.}\ \bibnamefont {Maan}}, \bibinfo {author}
  {\bibfnamefont {T.}~\bibnamefont {Rasing}}, \bibinfo {author} {\bibfnamefont
  {P.~C.~M.}\ \bibnamefont {Christianen}},\ and\ \bibinfo {author}
  {\bibfnamefont {A.~V.}\ \bibnamefont {Kimel}},\ }\bibfield  {title} {\bibinfo
  {title} {Ultrafast magnetism of a ferrimagnet across the spin-flop transition
  in high magnetic fields},\ }\href
  {https://doi.org/10.1103/PhysRevLett.118.117203} {\bibfield  {journal}
  {\bibinfo  {journal} {Phys. Rev. Lett.}\ }\textbf {\bibinfo {volume} {118}},\
  \bibinfo {pages} {117203} (\bibinfo {year} {2017})}\BibitemShut {NoStop}%
\bibitem [{\citenamefont {Hummler}\ and\ \citenamefont
  {F\"ahnle}(1996)}]{MolecularField_DyinCo}%
  \BibitemOpen
  \bibfield  {author} {\bibinfo {author} {\bibfnamefont {K.}~\bibnamefont
  {Hummler}}\ and\ \bibinfo {author} {\bibfnamefont {M.}~\bibnamefont
  {F\"ahnle}},\ }\bibfield  {title} {\bibinfo {title} {Full-potential
  linear-muffin-tin-orbital calculations of the magnetic properties of
  rare-earth--transition-metal intermetallics. i. description of the formalism
  and application to the series r${\mathrm{co}}_{5}$ (r=rare-earth atom)},\
  }\href {https://doi.org/10.1103/PhysRevB.53.3272} {\bibfield  {journal}
  {\bibinfo  {journal} {Phys. Rev. B}\ }\textbf {\bibinfo {volume} {53}},\
  \bibinfo {pages} {3272} (\bibinfo {year} {1996})}\BibitemShut {NoStop}%
\bibitem [{\citenamefont {St\"ohr}(1999)}]{Stohr_JM3review}%
  \BibitemOpen
  \bibfield  {author} {\bibinfo {author} {\bibfnamefont {J.}~\bibnamefont
  {St\"ohr}},\ }\href@noop {} {\bibfield  {journal} {\bibinfo  {journal}
  {Journal of Magnetism and Magnetic Materials}\ }\textbf {\bibinfo {volume}
  {200}},\ \bibinfo {pages} {470 } (\bibinfo {year} {1999})}\BibitemShut
  {NoStop}%
\bibitem [{\citenamefont {Collins}\ \emph {et~al.}(1995)\citenamefont
  {Collins}, \citenamefont {Laundy}, \citenamefont {Tang},\ and\ \citenamefont
  {van~der Laan}}]{XMCD_US}%
  \BibitemOpen
  \bibfield  {author} {\bibinfo {author} {\bibfnamefont {S.~P.}\ \bibnamefont
  {Collins}}, \bibinfo {author} {\bibfnamefont {D.}~\bibnamefont {Laundy}},
  \bibinfo {author} {\bibfnamefont {C.~C.}\ \bibnamefont {Tang}},\ and\
  \bibinfo {author} {\bibfnamefont {G.}~\bibnamefont {van~der Laan}},\
  }\bibfield  {title} {\bibinfo {title} {An investigation of uranium m 4,5 edge
  magnetic x-ray circular dichroism in us},\ }\href@noop {} {\bibfield
  {journal} {\bibinfo  {journal} {Journal of Physics: Condensed Matter}\
  }\textbf {\bibinfo {volume} {7}},\ \bibinfo {pages} {9325} (\bibinfo {year}
  {1995})}\BibitemShut {NoStop}%
\bibitem [{\citenamefont {Teramura}\ \emph {et~al.}(1996)\citenamefont
  {Teramura}, \citenamefont {Tanaka}, \citenamefont {Thole},\ and\
  \citenamefont {Jo}}]{teramura_RE}%
  \BibitemOpen
  \bibfield  {author} {\bibinfo {author} {\bibfnamefont {Y.}~\bibnamefont
  {Teramura}}, \bibinfo {author} {\bibfnamefont {A.}~\bibnamefont {Tanaka}},
  \bibinfo {author} {\bibfnamefont {B.}~\bibnamefont {Thole}},\ and\ \bibinfo
  {author} {\bibfnamefont {T.}~\bibnamefont {Jo}},\ }\href@noop {} {\bibfield
  {journal} {\bibinfo  {journal} {Journal of the Physical Society of Japan}\
  }\textbf {\bibinfo {volume} {65}},\ \bibinfo {pages} {3056} (\bibinfo {year}
  {1996})}\BibitemShut {NoStop}%
\bibitem [{\citenamefont {Bergeard}\ \emph {et~al.}(2017)\citenamefont
  {Bergeard}, \citenamefont {Mougin}, \citenamefont {Izquierdo}, \citenamefont
  {Fonda},\ and\ \citenamefont {Sirotti}}]{GdCo_segre_PRB}%
  \BibitemOpen
  \bibfield  {author} {\bibinfo {author} {\bibfnamefont {N.}~\bibnamefont
  {Bergeard}}, \bibinfo {author} {\bibfnamefont {A.}~\bibnamefont {Mougin}},
  \bibinfo {author} {\bibfnamefont {M.}~\bibnamefont {Izquierdo}}, \bibinfo
  {author} {\bibfnamefont {E.}~\bibnamefont {Fonda}},\ and\ \bibinfo {author}
  {\bibfnamefont {F.}~\bibnamefont {Sirotti}},\ }\bibfield  {title} {\bibinfo
  {title} {Correlation between structure, electronic properties, and magnetism
  in ${\mathrm{co}}_{x}{\mathrm{gd}}_{1\ensuremath{-}x}$ thin amorphous
  films},\ }\href {https://doi.org/10.1103/PhysRevB.96.064418} {\bibfield
  {journal} {\bibinfo  {journal} {Phys. Rev. B}\ }\textbf {\bibinfo {volume}
  {96}},\ \bibinfo {pages} {064418} (\bibinfo {year} {2017})}\BibitemShut
  {NoStop}%
\bibitem [{\citenamefont {Haltz}\ \emph {et~al.}(2018)\citenamefont {Haltz},
  \citenamefont {Weil}, \citenamefont {Sampaio}, \citenamefont {Pointillon},
  \citenamefont {Rousseau}, \citenamefont {March}, \citenamefont {Brun},
  \citenamefont {Li}, \citenamefont {Briand}, \citenamefont {Bachelet},
  \citenamefont {Dumont},\ and\ \citenamefont {Mougin}}]{TbFeCo_segre_PRM}%
  \BibitemOpen
  \bibfield  {author} {\bibinfo {author} {\bibfnamefont {E.}~\bibnamefont
  {Haltz}}, \bibinfo {author} {\bibfnamefont {R.}~\bibnamefont {Weil}},
  \bibinfo {author} {\bibfnamefont {J.}~\bibnamefont {Sampaio}}, \bibinfo
  {author} {\bibfnamefont {A.}~\bibnamefont {Pointillon}}, \bibinfo {author}
  {\bibfnamefont {O.}~\bibnamefont {Rousseau}}, \bibinfo {author}
  {\bibfnamefont {K.}~\bibnamefont {March}}, \bibinfo {author} {\bibfnamefont
  {N.}~\bibnamefont {Brun}}, \bibinfo {author} {\bibfnamefont {Z.}~\bibnamefont
  {Li}}, \bibinfo {author} {\bibfnamefont {E.}~\bibnamefont {Briand}}, \bibinfo
  {author} {\bibfnamefont {C.}~\bibnamefont {Bachelet}}, \bibinfo {author}
  {\bibfnamefont {Y.}~\bibnamefont {Dumont}},\ and\ \bibinfo {author}
  {\bibfnamefont {A.}~\bibnamefont {Mougin}},\ }\bibfield  {title} {\bibinfo
  {title} {Deviations from bulk behavior in tbfe(co) thin films: Interfaces
  contribution in the biased composition},\ }\href
  {https://doi.org/10.1103/PhysRevMaterials.2.104410} {\bibfield  {journal}
  {\bibinfo  {journal} {Phys. Rev. Materials}\ }\textbf {\bibinfo {volume}
  {2}},\ \bibinfo {pages} {104410} (\bibinfo {year} {2018})}\BibitemShut
  {NoStop}%
\bibitem [{\citenamefont {Baczewski}\ \emph {et~al.}(1993)\citenamefont
  {Baczewski}, \citenamefont {Givord}, \citenamefont {Alameda}, \citenamefont
  {Dieny}, \citenamefont {Nozieres}, \citenamefont {Rebouillat},\ and\
  \citenamefont {Prejean}}]{Givord}%
  \BibitemOpen
  \bibfield  {author} {\bibinfo {author} {\bibfnamefont {L.}~\bibnamefont
  {Baczewski}}, \bibinfo {author} {\bibfnamefont {D.}~\bibnamefont {Givord}},
  \bibinfo {author} {\bibfnamefont {J.}~\bibnamefont {Alameda}}, \bibinfo
  {author} {\bibfnamefont {B.}~\bibnamefont {Dieny}}, \bibinfo {author}
  {\bibfnamefont {J.}~\bibnamefont {Nozieres}}, \bibinfo {author}
  {\bibfnamefont {J.}~\bibnamefont {Rebouillat}},\ and\ \bibinfo {author}
  {\bibfnamefont {J.}~\bibnamefont {Prejean}},\ }\bibfield  {title} {\bibinfo
  {title} {Magnetism in rare-earth-transition metal systems. magnetization
  reversal and ultra-high susceptibility in sandwiched thin films based on
  rare-earth and cobalt alloys},\ }\href
  {https://doi.org/10.12693/APhysPolA.83.629} {\bibfield  {journal} {\bibinfo
  {journal} {ACTA PHYSICA POLONICA A}\ }\textbf {\bibinfo {volume} {83}},\
  \bibinfo {pages} {629} (\bibinfo {year} {1993})}\BibitemShut {NoStop}%
\bibitem [{\citenamefont {Luo}\ \emph {et~al.}(2019)\citenamefont {Luo},
  \citenamefont {Ryll}, \citenamefont {Back},\ and\ \citenamefont
  {Radu}}]{SciRep_2019}%
  \BibitemOpen
  \bibfield  {author} {\bibinfo {author} {\bibfnamefont {C.}~\bibnamefont
  {Luo}}, \bibinfo {author} {\bibfnamefont {H.}~\bibnamefont {Ryll}}, \bibinfo
  {author} {\bibfnamefont {C.}~\bibnamefont {Back}},\ and\ \bibinfo {author}
  {\bibfnamefont {F.}~\bibnamefont {Radu}},\ }\bibfield  {title} {\bibinfo
  {title} {X-ray magnetic linear dichroism as a probe for non-collinear
  magnetic state in ferrimagnetic single layer exchange bias systems},\ }\href
  {https://doi.org/10.1038/s41598-019-54356-y} {\bibfield  {journal} {\bibinfo
  {journal} {Scientific Reports}\ }\textbf {\bibinfo {volume} {9}},\ \bibinfo
  {pages} {18169} (\bibinfo {year} {2019})}\BibitemShut {NoStop}%
\bibitem [{\citenamefont {Cid}\ \emph {et~al.}(2017)\citenamefont {Cid},
  \citenamefont {Alameda}, \citenamefont {Valvidares}, \citenamefont {Cezar},
  \citenamefont {Bencok}, \citenamefont {Brookes},\ and\ \citenamefont
  {D\'{\i}az}}]{NdCo_PRB}%
  \BibitemOpen
  \bibfield  {author} {\bibinfo {author} {\bibfnamefont {R.}~\bibnamefont
  {Cid}}, \bibinfo {author} {\bibfnamefont {J.~M.}\ \bibnamefont {Alameda}},
  \bibinfo {author} {\bibfnamefont {S.~M.}\ \bibnamefont {Valvidares}},
  \bibinfo {author} {\bibfnamefont {J.~C.}\ \bibnamefont {Cezar}}, \bibinfo
  {author} {\bibfnamefont {P.}~\bibnamefont {Bencok}}, \bibinfo {author}
  {\bibfnamefont {N.~B.}\ \bibnamefont {Brookes}},\ and\ \bibinfo {author}
  {\bibfnamefont {J.}~\bibnamefont {D\'{\i}az}},\ }\bibfield  {title} {\bibinfo
  {title} {Perpendicular magnetic anisotropy in amorphous
  ${\mathrm{nd}}_{x}{\mathrm{co}}_{1\ensuremath{-}x}$ thin films studied by
  x-ray magnetic circular dichroism},\ }\href
  {https://doi.org/10.1103/PhysRevB.95.224402} {\bibfield  {journal} {\bibinfo
  {journal} {Phys. Rev. B}\ }\textbf {\bibinfo {volume} {95}},\ \bibinfo
  {pages} {224402} (\bibinfo {year} {2017})}\BibitemShut {NoStop}%
\bibitem [{\citenamefont {D{\'{\i}}az}\ \emph {et~al.}(2013)\citenamefont
  {D{\'{\i}}az}, \citenamefont {Cid}, \citenamefont {Hierro}, \citenamefont
  {{\'{A}}lvarez-Prado}, \citenamefont {Quir{\'{o}}s},\ and\ \citenamefont
  {Alameda}}]{NdCo_EXAFS}%
  \BibitemOpen
  \bibfield  {author} {\bibinfo {author} {\bibfnamefont {J.}~\bibnamefont
  {D{\'{\i}}az}}, \bibinfo {author} {\bibfnamefont {R.}~\bibnamefont {Cid}},
  \bibinfo {author} {\bibfnamefont {A.}~\bibnamefont {Hierro}}, \bibinfo
  {author} {\bibfnamefont {L.~M.}\ \bibnamefont {{\'{A}}lvarez-Prado}},
  \bibinfo {author} {\bibfnamefont {C.}~\bibnamefont {Quir{\'{o}}s}},\ and\
  \bibinfo {author} {\bibfnamefont {J.~M.}\ \bibnamefont {Alameda}},\
  }\bibfield  {title} {\bibinfo {title} {Large negative thermal expansion of
  the co subnetwork measured by {EXAFS} in highly disordered nd1-{xCoxthin}
  films with perpendicular magnetic anisotropy},\ }\href
  {https://doi.org/10.1088/0953-8984/25/42/426002} {\bibfield  {journal}
  {\bibinfo  {journal} {Journal of Physics: Condensed Matter}\ }\textbf
  {\bibinfo {volume} {25}},\ \bibinfo {pages} {426002} (\bibinfo {year}
  {2013})}\BibitemShut {NoStop}%
\bibitem [{\citenamefont {Davydova}\ \emph {et~al.}(2019)\citenamefont
  {Davydova}, \citenamefont {Zvezdin}, \citenamefont {Becker}, \citenamefont
  {Kimel},\ and\ \citenamefont {Zvezdin}}]{HT_phaseDiagram}%
  \BibitemOpen
  \bibfield  {author} {\bibinfo {author} {\bibfnamefont {M.~D.}\ \bibnamefont
  {Davydova}}, \bibinfo {author} {\bibfnamefont {K.~A.}\ \bibnamefont
  {Zvezdin}}, \bibinfo {author} {\bibfnamefont {J.}~\bibnamefont {Becker}},
  \bibinfo {author} {\bibfnamefont {A.~V.}\ \bibnamefont {Kimel}},\ and\
  \bibinfo {author} {\bibfnamefont {A.~K.}\ \bibnamefont {Zvezdin}},\
  }\bibfield  {title} {\bibinfo {title} {$h\text{\ensuremath{-}}t$ phase
  diagram of rare-earth--transition-metal alloys in the vicinity of the
  compensation point},\ }\href {https://doi.org/10.1103/PhysRevB.100.064409}
  {\bibfield  {journal} {\bibinfo  {journal} {Phys. Rev. B}\ }\textbf {\bibinfo
  {volume} {100}},\ \bibinfo {pages} {064409} (\bibinfo {year}
  {2019})}\BibitemShut {NoStop}%
\bibitem [{\citenamefont {Chen}\ \emph
  {et~al.}(2015{\natexlab{b}})\citenamefont {Chen}, \citenamefont {Lott},
  \citenamefont {Radu}, \citenamefont {Choueikani}, \citenamefont {Otero},\
  and\ \citenamefont {Ohresser}}]{deconvoluted_DyCo}%
  \BibitemOpen
  \bibfield  {author} {\bibinfo {author} {\bibfnamefont {K.}~\bibnamefont
  {Chen}}, \bibinfo {author} {\bibfnamefont {D.}~\bibnamefont {Lott}}, \bibinfo
  {author} {\bibfnamefont {F.}~\bibnamefont {Radu}}, \bibinfo {author}
  {\bibfnamefont {F.}~\bibnamefont {Choueikani}}, \bibinfo {author}
  {\bibfnamefont {E.}~\bibnamefont {Otero}},\ and\ \bibinfo {author}
  {\bibfnamefont {P.}~\bibnamefont {Ohresser}},\ }\bibfield  {title} {\bibinfo
  {title} {Temperature-dependent magnetic properties of ferrimagnetic
  ${\mathrm{dyco}}_{3}$ alloy films},\ }\href
  {https://doi.org/10.1103/PhysRevB.91.024409} {\bibfield  {journal} {\bibinfo
  {journal} {Phys. Rev. B}\ }\textbf {\bibinfo {volume} {91}},\ \bibinfo
  {pages} {024409} (\bibinfo {year} {2015}{\natexlab{b}})}\BibitemShut
  {NoStop}%
\bibitem [{\citenamefont {Barla}\ \emph {et~al.}(2016)\citenamefont {Barla},
  \citenamefont {Nicol{\'{a}}s}, \citenamefont {Cocco}, \citenamefont
  {Valvidares}, \citenamefont {Herrero-Mart{\'\i}n}, \citenamefont {Gargiani},
  \citenamefont {Moldes}, \citenamefont {Ruget}, \citenamefont {Pellegrin},\
  and\ \citenamefont {Ferrer}}]{HECTOR}%
  \BibitemOpen
  \bibfield  {author} {\bibinfo {author} {\bibfnamefont {A.}~\bibnamefont
  {Barla}}, \bibinfo {author} {\bibfnamefont {J.}~\bibnamefont
  {Nicol{\'{a}}s}}, \bibinfo {author} {\bibfnamefont {D.}~\bibnamefont
  {Cocco}}, \bibinfo {author} {\bibfnamefont {S.~M.}\ \bibnamefont
  {Valvidares}}, \bibinfo {author} {\bibfnamefont {J.}~\bibnamefont
  {Herrero-Mart{\'\i}n}}, \bibinfo {author} {\bibfnamefont {P.}~\bibnamefont
  {Gargiani}}, \bibinfo {author} {\bibfnamefont {J.}~\bibnamefont {Moldes}},
  \bibinfo {author} {\bibfnamefont {C.}~\bibnamefont {Ruget}}, \bibinfo
  {author} {\bibfnamefont {E.}~\bibnamefont {Pellegrin}},\ and\ \bibinfo
  {author} {\bibfnamefont {S.}~\bibnamefont {Ferrer}},\ }\bibfield  {title}
  {\bibinfo {title} {{Design and performance of BOREAS, the beamline for
  resonant X-ray absorption and scattering experiments at the ALBA synchrotron
  light source}},\ }\href {https://doi.org/10.1107/S1600577516013461}
  {\bibfield  {journal} {\bibinfo  {journal} {Journal of Synchrotron
  Radiation}\ }\textbf {\bibinfo {volume} {23}},\ \bibinfo {pages} {1507}
  (\bibinfo {year} {2016})}\BibitemShut {NoStop}%
\bibitem [{\citenamefont {Blanco-Rold\'an}(2017)}]{Tesis_Cristina}%
  \BibitemOpen
  \bibfield  {author} {\bibinfo {author} {\bibfnamefont {C.}~\bibnamefont
  {Blanco-Rold\'an}},\ }\href {http://hdl.handle.net/10651/44970} {\bibinfo
  {title} {Magnetic interactions in rare earth-transition metal systems and
  their study by synchrotron radiation techniques}} (\bibinfo {year}
  {2017})\BibitemShut {NoStop}%
\bibitem [{\citenamefont {Altarelli}(1993)}]{XMCD_L_SumRule_Altarelli}%
  \BibitemOpen
  \bibfield  {author} {\bibinfo {author} {\bibfnamefont {M.}~\bibnamefont
  {Altarelli}},\ }\bibfield  {title} {\bibinfo {title} {Orbital-magnetization
  sum rule for x-ray circular dichroism: A simple proof},\ }\href@noop {}
  {\bibfield  {journal} {\bibinfo  {journal} {Phys. Rev. B}\ }\textbf {\bibinfo
  {volume} {47}},\ \bibinfo {pages} {597} (\bibinfo {year} {1993})}\BibitemShut
  {NoStop}%
\bibitem [{\citenamefont {Carra}\ \emph {et~al.}(1993)\citenamefont {Carra},
  \citenamefont {Thole}, \citenamefont {Altarelli},\ and\ \citenamefont
  {Wang}}]{Carra_PRL}%
  \BibitemOpen
  \bibfield  {author} {\bibinfo {author} {\bibfnamefont {P.}~\bibnamefont
  {Carra}}, \bibinfo {author} {\bibfnamefont {B.~T.}\ \bibnamefont {Thole}},
  \bibinfo {author} {\bibfnamefont {M.}~\bibnamefont {Altarelli}},\ and\
  \bibinfo {author} {\bibfnamefont {X.}~\bibnamefont {Wang}},\ }\bibfield
  {title} {\bibinfo {title} {X-ray circular dichroism and local magnetic
  fields},\ }\href@noop {} {\bibfield  {journal} {\bibinfo  {journal} {Phys.
  Rev. Lett.}\ }\textbf {\bibinfo {volume} {70}},\ \bibinfo {pages} {694}
  (\bibinfo {year} {1993})}\BibitemShut {NoStop}%
\bibitem [{\citenamefont {Chen}\ \emph {et~al.}(1995)\citenamefont {Chen},
  \citenamefont {Idzerda}, \citenamefont {Lin}, \citenamefont {Smith},
  \citenamefont {Meigs}, \citenamefont {Chaban}, \citenamefont {Ho},
  \citenamefont {Pellegrin},\ and\ \citenamefont {Sette}}]{Co_XMCD_ChenPRL}%
  \BibitemOpen
  \bibfield  {author} {\bibinfo {author} {\bibfnamefont {C.~T.}\ \bibnamefont
  {Chen}}, \bibinfo {author} {\bibfnamefont {Y.~U.}\ \bibnamefont {Idzerda}},
  \bibinfo {author} {\bibfnamefont {H.-J.}\ \bibnamefont {Lin}}, \bibinfo
  {author} {\bibfnamefont {N.~V.}\ \bibnamefont {Smith}}, \bibinfo {author}
  {\bibfnamefont {G.}~\bibnamefont {Meigs}}, \bibinfo {author} {\bibfnamefont
  {E.}~\bibnamefont {Chaban}}, \bibinfo {author} {\bibfnamefont {G.~H.}\
  \bibnamefont {Ho}}, \bibinfo {author} {\bibfnamefont {E.}~\bibnamefont
  {Pellegrin}},\ and\ \bibinfo {author} {\bibfnamefont {F.}~\bibnamefont
  {Sette}},\ }\bibfield  {title} {\bibinfo {title} {Experimental confirmation
  of the x-ray magnetic circular dichroism sum rules for iron and cobalt},\
  }\href {https://doi.org/10.1103/PhysRevLett.75.152} {\bibfield  {journal}
  {\bibinfo  {journal} {Phys. Rev. Lett.}\ }\textbf {\bibinfo {volume} {75}},\
  \bibinfo {pages} {152} (\bibinfo {year} {1995})}\BibitemShut {NoStop}%
\bibitem [{\citenamefont {St{\"o}hr}\ and\ \citenamefont
  {K{\"o}nig}(1995)}]{Stohr_XMCDAngle}%
  \BibitemOpen
  \bibfield  {author} {\bibinfo {author} {\bibfnamefont {J.}~\bibnamefont
  {St{\"o}hr}}\ and\ \bibinfo {author} {\bibfnamefont {H.}~\bibnamefont
  {K{\"o}nig}},\ }\bibfield  {title} {\bibinfo {title} {Determination of
  spin-and orbital-moment anisotropies in transition metals by angle-dependent
  x-ray magnetic circular dichroism},\ }\href@noop {} {\bibfield  {journal}
  {\bibinfo  {journal} {Physical review letters}\ }\textbf {\bibinfo {volume}
  {75}},\ \bibinfo {pages} {3748} (\bibinfo {year} {1995})}\BibitemShut
  {NoStop}%
\bibitem [{\citenamefont {Van Der~Laan}(1998)}]{vanderlaan}%
  \BibitemOpen
  \bibfield  {author} {\bibinfo {author} {\bibfnamefont {G.}~\bibnamefont {Van
  Der~Laan}},\ }\bibfield  {title} {\bibinfo {title} {Microscopic origin of
  magnetocrystalline anisotropy in transition metal thin films},\ }\href
  {https://doi.org/10.1088/0953-8984/10/14/012} {\bibfield  {journal} {\bibinfo
   {journal} {Journal of Physics Condensed Matter}\ }\textbf {\bibinfo {volume}
  {10}},\ \bibinfo {pages} {3239} (\bibinfo {year} {1998})}\BibitemShut
  {NoStop}%
\bibitem [{\citenamefont {Suzuki}\ and\ \citenamefont
  {Miwa}(2019)}]{theory_japos2019}%
  \BibitemOpen
  \bibfield  {author} {\bibinfo {author} {\bibfnamefont {Y.}~\bibnamefont
  {Suzuki}}\ and\ \bibinfo {author} {\bibfnamefont {S.}~\bibnamefont {Miwa}},\
  }\bibfield  {title} {\bibinfo {title} {Magnetic anisotropy of ferromagnetic
  metals in low-symmetry systems},\ }\href
  {https://doi.org/https://doi.org/10.1016/j.physleta.2019.01.020} {\bibfield
  {journal} {\bibinfo  {journal} {Physics Letters A}\ }\textbf {\bibinfo
  {volume} {383}},\ \bibinfo {pages} {1203 } (\bibinfo {year}
  {2019})}\BibitemShut {NoStop}%
\bibitem [{\citenamefont {Weller}\ \emph {et~al.}(1995)\citenamefont {Weller},
  \citenamefont {St\"ohr}, \citenamefont {Nakajima}, \citenamefont {Carl},
  \citenamefont {Samant}, \citenamefont {Chappert}, \citenamefont {M\'egy},
  \citenamefont {Beauvillain}, \citenamefont {Veillet},\ and\ \citenamefont
  {Held}}]{Stohr_CoAu}%
  \BibitemOpen
  \bibfield  {author} {\bibinfo {author} {\bibfnamefont {D.}~\bibnamefont
  {Weller}}, \bibinfo {author} {\bibfnamefont {J.}~\bibnamefont {St\"ohr}},
  \bibinfo {author} {\bibfnamefont {R.}~\bibnamefont {Nakajima}}, \bibinfo
  {author} {\bibfnamefont {A.}~\bibnamefont {Carl}}, \bibinfo {author}
  {\bibfnamefont {M.~G.}\ \bibnamefont {Samant}}, \bibinfo {author}
  {\bibfnamefont {C.}~\bibnamefont {Chappert}}, \bibinfo {author}
  {\bibfnamefont {R.}~\bibnamefont {M\'egy}}, \bibinfo {author} {\bibfnamefont
  {P.}~\bibnamefont {Beauvillain}}, \bibinfo {author} {\bibfnamefont
  {P.}~\bibnamefont {Veillet}},\ and\ \bibinfo {author} {\bibfnamefont {G.~A.}\
  \bibnamefont {Held}},\ }\bibfield  {title} {\bibinfo {title} {Microscopic
  origin of magnetic anisotropy in au/co/au probed with x-ray magnetic circular
  dichroism},\ }\href@noop {} {\bibfield  {journal} {\bibinfo  {journal} {Phys.
  Rev. Lett.}\ }\textbf {\bibinfo {volume} {75}},\ \bibinfo {pages} {3752}
  (\bibinfo {year} {1995})}\BibitemShut {NoStop}%
\bibitem [{\citenamefont {Wu}\ and\ \citenamefont
  {Chuang}(1991)}]{YCo_PhaseDiagram}%
  \BibitemOpen
  \bibfield  {author} {\bibinfo {author} {\bibfnamefont {C.~H.}\ \bibnamefont
  {Wu}}\ and\ \bibinfo {author} {\bibfnamefont {Y.~C.}\ \bibnamefont
  {Chuang}},\ }\bibfield  {title} {\bibinfo {title} {The co-y (cobalt-yttrium)
  system},\ }\href {https://doi.org/10.1007/BF02645075} {\bibfield  {journal}
  {\bibinfo  {journal} {Journal of Phase Equilibria}\ }\textbf {\bibinfo
  {volume} {12}},\ \bibinfo {pages} {587} (\bibinfo {year} {1991})}\BibitemShut
  {NoStop}%
\bibitem [{\citenamefont {Predel}(2012)}]{DyCo_PhaseDiagram}%
  \BibitemOpen
  \bibfield  {author} {\bibinfo {author} {\bibfnamefont {B.}~\bibnamefont
  {Predel}},\ }\href {https://doi.org/10.1007/978-3-540-44756-6_157} {\bibinfo
  {title} {Co - dy (cobalt - dysprosium): Datasheet from landolt-b{\"o}rnstein
  - group iv physical chemistry}} (\bibinfo {year} {2012})\BibitemShut
  {NoStop}%
\bibitem [{\citenamefont {Coey}(1985)}]{amorphous_magnets_Coey}%
  \BibitemOpen
  \bibfield  {author} {\bibinfo {author} {\bibfnamefont {J.~M.~D.}\
  \bibnamefont {Coey}},\ }\bibinfo {title} {Magnetism in amorphous solids},\
  in\ \href {https://doi.org/10.1007/978-1-4757-9156-3_13} {\emph {\bibinfo
  {booktitle} {Amorphous Solids and the Liquid State}}},\ \bibinfo {editor}
  {edited by\ \bibinfo {editor} {\bibfnamefont {N.~H.}\ \bibnamefont {March}},
  \bibinfo {editor} {\bibfnamefont {R.~A.}\ \bibnamefont {Street}},\ and\
  \bibinfo {editor} {\bibfnamefont {M.~P.}\ \bibnamefont {Tosi}}}\ (\bibinfo
  {publisher} {Springer US},\ \bibinfo {address} {Boston, MA},\ \bibinfo {year}
  {1985})\ pp.\ \bibinfo {pages} {433--466}\BibitemShut {NoStop}%
\bibitem [{\citenamefont {Patrick}\ and\ \citenamefont
  {Staunton}(2018)}]{2018_Tc_RETM}%
  \BibitemOpen
  \bibfield  {author} {\bibinfo {author} {\bibfnamefont {C.~E.}\ \bibnamefont
  {Patrick}}\ and\ \bibinfo {author} {\bibfnamefont {J.~B.}\ \bibnamefont
  {Staunton}},\ }\bibfield  {title} {\bibinfo {title}
  {Rare-earth/transition-metal magnets at finite temperature:
  Self-interaction-corrected relativistic density functional theory in the
  disordered local moment picture},\ }\href
  {https://doi.org/10.1103/PhysRevB.97.224415} {\bibfield  {journal} {\bibinfo
  {journal} {Phys. Rev. B}\ }\textbf {\bibinfo {volume} {97}},\ \bibinfo
  {pages} {224415} (\bibinfo {year} {2018})}\BibitemShut {NoStop}%
\bibitem [{\citenamefont {Thole}\ \emph {et~al.}(1985)\citenamefont {Thole},
  \citenamefont {van~der Laan}, \citenamefont {Fuggle}, \citenamefont
  {Sawatzky}, \citenamefont {Karnatak},\ and\ \citenamefont
  {Esteva}}]{Thole_PRB}%
  \BibitemOpen
  \bibfield  {author} {\bibinfo {author} {\bibfnamefont {B.~T.}\ \bibnamefont
  {Thole}}, \bibinfo {author} {\bibfnamefont {G.}~\bibnamefont {van~der Laan}},
  \bibinfo {author} {\bibfnamefont {J.~C.}\ \bibnamefont {Fuggle}}, \bibinfo
  {author} {\bibfnamefont {G.~A.}\ \bibnamefont {Sawatzky}}, \bibinfo {author}
  {\bibfnamefont {R.~C.}\ \bibnamefont {Karnatak}},\ and\ \bibinfo {author}
  {\bibfnamefont {J.-M.}\ \bibnamefont {Esteva}},\ }\bibfield  {title}
  {\bibinfo {title} {3d x-ray-absorption lines and the 3${d}^{9}$4${f}^{n+1}$
  multiplets of the lanthanides},\ }\href
  {https://doi.org/10.1103/PhysRevB.32.5107} {\bibfield  {journal} {\bibinfo
  {journal} {Phys. Rev. B}\ }\textbf {\bibinfo {volume} {32}},\ \bibinfo
  {pages} {5107} (\bibinfo {year} {1985})}\BibitemShut {NoStop}%
\bibitem [{\citenamefont {Goedkoop}(1985)}]{Goedkoop_thesis}%
  \BibitemOpen
  \bibfield  {author} {\bibinfo {author} {\bibfnamefont {J.}~\bibnamefont
  {Goedkoop}},\ }\href
  {http://inis.iaea.org/search/search.aspx?orig_q=RN:21039034} {\bibinfo
  {title} {X-ray dichroism of rare earth materials}} (\bibinfo {year}
  {1985})\BibitemShut {NoStop}%
\bibitem [{\citenamefont {Goedkoop}\ \emph {et~al.}(1988)\citenamefont
  {Goedkoop}, \citenamefont {Thole}, \citenamefont {van~der Laan},
  \citenamefont {Sawatzky}, \citenamefont {de~Groot},\ and\ \citenamefont
  {Fuggle}}]{Goedkoop_PRB}%
  \BibitemOpen
  \bibfield  {author} {\bibinfo {author} {\bibfnamefont {J.~B.}\ \bibnamefont
  {Goedkoop}}, \bibinfo {author} {\bibfnamefont {B.~T.}\ \bibnamefont {Thole}},
  \bibinfo {author} {\bibfnamefont {G.}~\bibnamefont {van~der Laan}}, \bibinfo
  {author} {\bibfnamefont {G.~A.}\ \bibnamefont {Sawatzky}}, \bibinfo {author}
  {\bibfnamefont {F.~M.~F.}\ \bibnamefont {de~Groot}},\ and\ \bibinfo {author}
  {\bibfnamefont {J.~C.}\ \bibnamefont {Fuggle}},\ }\bibfield  {title}
  {\bibinfo {title} {Calculations of magnetic x-ray dichroism in the 3d
  absorption spectra of rare-earth compounds},\ }\href
  {https://doi.org/10.1103/PhysRevB.37.2086} {\bibfield  {journal} {\bibinfo
  {journal} {Phys. Rev. B}\ }\textbf {\bibinfo {volume} {37}},\ \bibinfo
  {pages} {2086} (\bibinfo {year} {1988})}\BibitemShut {NoStop}%
\bibitem [{\citenamefont {van~der Laan}\ \emph {et~al.}(2008)\citenamefont
  {van~der Laan}, \citenamefont {Arenholz}, \citenamefont {Schmehl},\ and\
  \citenamefont {Schlom}}]{EuO_PRL}%
  \BibitemOpen
  \bibfield  {author} {\bibinfo {author} {\bibfnamefont {G.}~\bibnamefont
  {van~der Laan}}, \bibinfo {author} {\bibfnamefont {E.}~\bibnamefont
  {Arenholz}}, \bibinfo {author} {\bibfnamefont {A.}~\bibnamefont {Schmehl}},\
  and\ \bibinfo {author} {\bibfnamefont {D.~G.}\ \bibnamefont {Schlom}},\
  }\bibfield  {title} {\bibinfo {title} {Weak anisotropic x-ray magnetic linear
  dichroism at the eu ${M}_{4,5}$ edges of ferromagnetic euo(001): Evidence for
  $4f$-state contributions},\ }\href@noop {} {\bibfield  {journal} {\bibinfo
  {journal} {Phys. Rev. Lett.}\ }\textbf {\bibinfo {volume} {100}},\ \bibinfo
  {pages} {067403} (\bibinfo {year} {2008})}\BibitemShut {NoStop}%
\bibitem [{\citenamefont {Das}\ \emph {et~al.}(2019)\citenamefont {Das},
  \citenamefont {Choudhary}, \citenamefont {Skomski}, \citenamefont
  {Balasubramanian}, \citenamefont {Pathak}, \citenamefont {Paudyal},\ and\
  \citenamefont {Sellmyer}}]{SmCo_anisotropy}%
  \BibitemOpen
  \bibfield  {author} {\bibinfo {author} {\bibfnamefont {B.}~\bibnamefont
  {Das}}, \bibinfo {author} {\bibfnamefont {R.}~\bibnamefont {Choudhary}},
  \bibinfo {author} {\bibfnamefont {R.}~\bibnamefont {Skomski}}, \bibinfo
  {author} {\bibfnamefont {B.}~\bibnamefont {Balasubramanian}}, \bibinfo
  {author} {\bibfnamefont {A.~K.}\ \bibnamefont {Pathak}}, \bibinfo {author}
  {\bibfnamefont {D.}~\bibnamefont {Paudyal}},\ and\ \bibinfo {author}
  {\bibfnamefont {D.~J.}\ \bibnamefont {Sellmyer}},\ }\bibfield  {title}
  {\bibinfo {title} {Anisotropy and orbital moment in sm-co permanent
  magnets},\ }\href {https://doi.org/10.1103/PhysRevB.100.024419} {\bibfield
  {journal} {\bibinfo  {journal} {Phys. Rev. B}\ }\textbf {\bibinfo {volume}
  {100}},\ \bibinfo {pages} {024419} (\bibinfo {year} {2019})}\BibitemShut
  {NoStop}%
\bibitem [{\citenamefont {Nakajima}\ \emph {et~al.}(1999)\citenamefont
  {Nakajima}, \citenamefont {St\"ohr},\ and\ \citenamefont
  {Idzerda}}]{saturation_stohr}%
  \BibitemOpen
  \bibfield  {author} {\bibinfo {author} {\bibfnamefont {R.}~\bibnamefont
  {Nakajima}}, \bibinfo {author} {\bibfnamefont {J.}~\bibnamefont {St\"ohr}},\
  and\ \bibinfo {author} {\bibfnamefont {Y.~U.}\ \bibnamefont {Idzerda}},\
  }\href@noop {} {\bibfield  {journal} {\bibinfo  {journal} {Phys. Rev. B}\
  }\textbf {\bibinfo {volume} {59}},\ \bibinfo {pages} {6421} (\bibinfo {year}
  {1999})}\BibitemShut {NoStop}%
\bibitem [{\citenamefont {Shinde}\ \emph {et~al.}(2020)\citenamefont {Shinde},
  \citenamefont {Tien}, \citenamefont {Huang}, \citenamefont {Park},
  \citenamefont {Yu}, \citenamefont {Chung},\ and\ \citenamefont
  {Kim}}]{AF_DyO}%
  \BibitemOpen
  \bibfield  {author} {\bibinfo {author} {\bibfnamefont {K.~P.}\ \bibnamefont
  {Shinde}}, \bibinfo {author} {\bibfnamefont {V.~M.}\ \bibnamefont {Tien}},
  \bibinfo {author} {\bibfnamefont {L.}~\bibnamefont {Huang}}, \bibinfo
  {author} {\bibfnamefont {H.-R.}\ \bibnamefont {Park}}, \bibinfo {author}
  {\bibfnamefont {S.-C.}\ \bibnamefont {Yu}}, \bibinfo {author} {\bibfnamefont
  {K.~C.}\ \bibnamefont {Chung}},\ and\ \bibinfo {author} {\bibfnamefont
  {D.-H.}\ \bibnamefont {Kim}},\ }\bibfield  {title} {\bibinfo {title}
  {Magnetocaloric effect in tb2o3 and dy2o3 nanoparticles at cryogenic
  temperatures},\ }\href {https://doi.org/10.1063/1.5120350} {\bibfield
  {journal} {\bibinfo  {journal} {Journal of Applied Physics}\ }\textbf
  {\bibinfo {volume} {127}},\ \bibinfo {pages} {054903} (\bibinfo {year}
  {2020})}\BibitemShut {NoStop}%
\bibitem [{\citenamefont {Anderson}\ \emph {et~al.}(2017)\citenamefont
  {Anderson}, \citenamefont {Zhang}, \citenamefont {Hupalo}, \citenamefont
  {Rosenberg}, \citenamefont {Freeland}, \citenamefont {Tringides},\ and\
  \citenamefont {Vaknin}}]{DyO_XAS}%
  \BibitemOpen
  \bibfield  {author} {\bibinfo {author} {\bibfnamefont {N.~A.}\ \bibnamefont
  {Anderson}}, \bibinfo {author} {\bibfnamefont {Q.}~\bibnamefont {Zhang}},
  \bibinfo {author} {\bibfnamefont {M.}~\bibnamefont {Hupalo}}, \bibinfo
  {author} {\bibfnamefont {R.~A.}\ \bibnamefont {Rosenberg}}, \bibinfo {author}
  {\bibfnamefont {J.~W.}\ \bibnamefont {Freeland}}, \bibinfo {author}
  {\bibfnamefont {M.~C.}\ \bibnamefont {Tringides}},\ and\ \bibinfo {author}
  {\bibfnamefont {D.}~\bibnamefont {Vaknin}},\ }\bibfield  {title} {\bibinfo
  {title} {Magnetic properties of dy nano-islands on graphene},\ }\href
  {https://doi.org/https://doi.org/10.1016/j.jmmm.2017.04.007} {\bibfield
  {journal} {\bibinfo  {journal} {Journal of Magnetism and Magnetic Materials}\
  }\textbf {\bibinfo {volume} {435}},\ \bibinfo {pages} {212 } (\bibinfo {year}
  {2017})}\BibitemShut {NoStop}%
\bibitem [{\citenamefont {Iglesias}\ and\ \citenamefont
  {Rubio}(2002)}]{Honorino_APL}%
  \BibitemOpen
  \bibfield  {author} {\bibinfo {author} {\bibfnamefont {R.}~\bibnamefont
  {Iglesias}}\ and\ \bibinfo {author} {\bibfnamefont {H.}~\bibnamefont
  {Rubio}},\ }\bibfield  {title} {\bibinfo {title} {Approach to saturation in
  nanomagnetic systems},\ }\href {https://doi.org/10.1063/1.1522835} {\bibfield
   {journal} {\bibinfo  {journal} {Journal of Applied Physics}\ }\textbf
  {\bibinfo {volume} {92}},\ \bibinfo {pages} {7696} (\bibinfo {year}
  {2002})}\BibitemShut {NoStop}%
\bibitem [{\citenamefont {Tonnerre}\ \emph {et~al.}(2008)\citenamefont
  {Tonnerre}, \citenamefont {De~Santis}, \citenamefont {Grenier}, \citenamefont
  {Tolentino}, \citenamefont {Langlais}, \citenamefont {Bontempi},
  \citenamefont {Garc\'{\i}a-Fern\'andez},\ and\ \citenamefont
  {Staub}}]{reflectivity_AF_F}%
  \BibitemOpen
  \bibfield  {author} {\bibinfo {author} {\bibfnamefont {J.~M.}\ \bibnamefont
  {Tonnerre}}, \bibinfo {author} {\bibfnamefont {M.}~\bibnamefont {De~Santis}},
  \bibinfo {author} {\bibfnamefont {S.}~\bibnamefont {Grenier}}, \bibinfo
  {author} {\bibfnamefont {H.~C.~N.}\ \bibnamefont {Tolentino}}, \bibinfo
  {author} {\bibfnamefont {V.}~\bibnamefont {Langlais}}, \bibinfo {author}
  {\bibfnamefont {E.}~\bibnamefont {Bontempi}}, \bibinfo {author}
  {\bibfnamefont {M.}~\bibnamefont {Garc\'{\i}a-Fern\'andez}},\ and\ \bibinfo
  {author} {\bibfnamefont {U.}~\bibnamefont {Staub}},\ }\bibfield  {title}
  {\bibinfo {title} {Depth magnetization profile of a perpendicular exchange
  coupled system by soft-x-ray resonant magnetic reflectivity},\ }\href
  {https://doi.org/10.1103/PhysRevLett.100.157202} {\bibfield  {journal}
  {\bibinfo  {journal} {Phys. Rev. Lett.}\ }\textbf {\bibinfo {volume} {100}},\
  \bibinfo {pages} {157202} (\bibinfo {year} {2008})}\BibitemShut {NoStop}%
\end{thebibliography}%

\end{document}